\newcommand{\tr}[1]{\mathrm{Tr}\left[ {#1} \right]} 

\newcommand{\W}{\mathcal{W}}
\newcommand{\D}{\mathcal{D}}
\newcommand{\I}{\mathcal{I}} 
\newcommand{\Arev}{A^{\text{rev}}}
\newcommand{\Airr}{A^{\text{irr}}} 
\newcommand{\Jrev}{J^{\text{rev}}}
\newcommand{\Jirr}{J^{\text{irr}}} 

\documentclass[twocolumn,superscriptaddress]{revtex4-1}
\usepackage{bm,natbib}
\usepackage{graphicx}
\usepackage{mathrsfs}
\usepackage{dcolumn,fancyhdr}
\usepackage{amsmath}
\usepackage{amssymb}
\usepackage{amsfonts,gensymb}
\usepackage{indentfirst}
\usepackage{bbold}
\usepackage{multirow}
\usepackage{dsfont}

\begin{document}
\title{Irreversibility and correlations in coupled quantum oscillators}
\author{M. Brunelli}
\affiliation{Centre for Theoretical Atomic, Molecular and Optical Physics, School of Mathematics and Physics, Queen's University, Belfast BT7 1NN, United Kingdom}
\author{M. Paternostro}
\affiliation{Centre for Theoretical Atomic, Molecular and Optical Physics, School of Mathematics and Physics, Queen's University, Belfast BT7 1NN, United Kingdom}

\begin{abstract}

We investigate the link between the irreversibility generated by a stationary dissipative process and the correlations established within a composite quantum system. 
We provide two equivalent expressions for the entropy generated in the non-equilibrium steady state of coupled quantum harmonic oscillators
that allow for a simple interpretation of the onset of irreversibility.
We then unveil a quantitative relation between the entropy production rate and correlations, both total and quantum, built between the two oscillators.  
In the small-coupling limit, the entropy production rate is shown to be proportional to both the mutual information and the quantum correlations.
We apply our results to the analysis of the optomechanical interaction between a nano-mechanical resonator and a cavity field, and show that
the behavior of the entropy production captures the peculiar features of the model, such as sideband cooling and amplification.

\end{abstract} 
\maketitle

The quantum-limited control of systems trespassing the microscopic domain requires a quantitative assessment of the irreversibility 
generated during their evolution. Realistic transformations in composite quantum systems take place at a finite rate, thus departing from the quasi-static limit, and unavoidably 
involve the interaction with the surroundings. Both these features bring about irreversibility, which is typically manifested by the generation of entropy 
and the dissipation of heat into the environment~\cite{IrrEnt}. However, despite being of primary importance, entropy production remains an elusive quantity, often inaccessible 
to direct probing. This is a particularly delicate point when applying thermodynamic considerations. Indeed, failing to properly take into account all the mechanisms responsible 
for dissipation (not known {\it a priori}) determines an underestimation of the entropy production~\cite{Esposito1}. This can lead to erroneous conclusions, for instance in the characterization of the 
efficiency of recently proposed quantum thermal machines and engines~\cite{KosloffRev}.
\par
Another defining feature of composite systems is that correlations may be shared among their constituents. 
In real-life scenarios, correlations are typically established as the result of some dynamical process which modifies the energetic and entropic balance of the system.  The building-up of correlations 
within a quantum system must then be accompanied by the production of entropy. 
In turn, this implies that irreversibility and correlations cannot be tackled separately. 

In this paper, we carry out a detailed study of the irreversibility generated in the stationary state of a bipartite quantum system. We consider a simple system
that yet encapsulates most of the salient features encountered in quantum-controlled experiments, consisting of two linearly interacting quantum oscillators, 
each dissipating into a local bath. We derive two equivalent expressions for the rate of irreversible entropy production at the steady state, linking the irreversibility 
to either the number of excitations stored in the two modes (with respect to their equilibrium values) or the correlations between their motion, thus providing a 
clear picture of the onset of irreversibility.  
\par
At variance with standard approaches, we investigate the entropic cost of establishing correlations in a realistic finite-time process. We compare the behavior 
of the entropy production to the total correlations and the quantum ones, quantified by the mutual information and the discord, respectively~\cite{VlatkoDiscord,
VlatkoRev}. We find that the former behaves monotonically with respect to the latter quantities. In particular, when the coupling between the systems is small compared to 
their natural frequencies, we find simple proportionality relations between the entropy production and either the mutual information or the quantum discord. In 
Ref.~\cite{CostCorr} the thermodynamic cost of establishing correlations has been addressed from a purely information-theoretical point of view, without 
addressing their dynamical origin. On the contrary, our analysis shows explicitly that generation of correlations and production of entropy are complementary aspects in 
a dissipative process. We apply our results to the characterization of irreversibility in an optomechaincal system~\cite{Aspelmeyer}. We show how the behavior 
of the entropy production fully accounts for the salient features of the system in different regimes, e.g. cavity-mediated cooling and amplification of the mechanical 
resonator.    
\par
Our results suggest that irreversibility and correlations are two sides of the same coin: 
as expected, generation of correlations, both classical and quantum, comes at the price of increasing the entropy of the system plus environment. Likewise, 
irreversibility in composite quantum systems, very much like correlations, can be viewed as a resource. Harness irreversible dynamics in a controllable way 
can be beneficial for quantum information processing, as we quantitatively assess.
\par
The remainder of this work is organized as follows: In Sec.~\ref{s:IrrOsc} we introduce the model and derive the expressions of the entropy production upon which 
our analysis relies. Sec.~\ref{s:EntProd} discusses in detail the features of the entropy production. In Sec.~\ref{s:EntCorr} we 
first define correlation measures based on the R\'enyi-2 entropy, which are instrumental to highlight the link between entropy production and correlations, and then 
present a detailed analysis of such a connection. In Sec.~\ref{s:OptoCorr}, our results are specialized to the relevant case of an optomechanical system. 
Sec.~\ref{s:Conclusions}  draws our conclusions and sketches the questions that are left open by our study. Finally, a set of four Appendices presents the most 
technical parts of our work.

\section{Irreversibility in coupled quantum oscillators}\label{s:IrrOsc}
The second law of thermodynamics, when referred to a system that exchanges energy with its surroundings, takes the familiar form~\cite{Callen}   
\begin{equation}\label{2ndLaw}
\Delta S\ge \int\frac{\delta Q}{T} \, ,
\end{equation}
where $\delta Q$ is the infinitesimal heat absorbed by the system. The strict inequality characterizes irreversible processes for which 
some energy is dissipated in the form of heat and lost into the environment~\cite{Prigogine}. 
Eq.~\eqref{2ndLaw} can be cast into an equality by highlighting the nonnegative mismatch and, for convenience, expressed in terms of rates, so that one finally has 
\begin{equation}\label{RateEq}
\frac{\text{d}S}{\text{d} t} =\Phi(t)+\Pi(t) \, ,
\end{equation}
where $\Pi(t)$ is the {\it irreversible entropy production rate} and $\Phi(t)$ is the entropy flux from the environment into the system. When the system attains a stationary state, such quantities take values $\Pi_{\text{s}}$ and $\Phi_{\text{s}}$ respectively, so that $\Pi_{\text{s}} = - \Phi_{\text{s}} > 0$, while only when both terms vanish one recovers
thermal equilibrium.
The entropy production rate thus accounts for the irreversibility of any physical process and represents 
a key quantity for the characterization of finite-time transformations and the performances of thermal machines. 
\par
Several characterizations of entropy production have been given, especially in the framework of stochastic thermodynamics. In such a context it has been shown that the 
entropy production rate  originates from the temporal-symmetry breaking at the level of single trajectories~\cite{Esposito1,Esposito2,EntProdArrow} and satisfies a 
fluctuation theorem~\cite{Seifert}, from which one can deduce its average non-negativity. In some limit, these conclusions have also been extended to open quantum 
systems~\cite{SpohnLeb, EntDeffner}. Interestingly, in Ref.~\cite{Esposito3} a characterization of entropy production as correlation between a system and a reservoir 
has been put forward, which is somehow close in spirit to our investigation. However, although entropy production is formally well characterized, given its full dependence 
on the microscopic trajectory of the system, few useful expressions for cases at hand are available and it remains an elusive quantity to measure, with just a handful of 
experimental measurements reported in literature~\cite{ExpSeifert,ExpNovotny,ExpPekola}. Recently a first measurement of the entropy production rate in driven-dissipative 
quantum systems has been reported~\cite{OurExp}. The rate of irreversible entropy production, as specified by Eq.~\eqref{RateEq}, is the main tool of our investigation, 
which we exploit to assess the irreversibility in the open dynamics of a bipartite quantum system. 
\par
We consider two quantum oscillators described by the field operators $\hat a$ and $\hat b$, having frequencies $\omega_{a, b}$ and masses $m_{a, b}$. 
The two oscillators are linearly coupled with strength $G$, so that the interacting system is described by  
\begin{equation}\label{Hamiltonian}
\hat{H}=\frac{\hbar \omega_a}{2} (\hat{q}_a^{2}+\hat{p}_a^{2})+\frac{\hbar \omega_b}{2} (\hat{q}_b^{2}+\hat{p}_b^{2})+\hbar G\hat{q}_a
\hat{q}_b,
\end{equation}
where $\hat{q}_{a,b}$ and $\hat{p}_{a,b}$  are dimensionless position and momentum operators. The system is also interacting with its surroundings, which we model as two 
independent Markovian baths at temperature $T_{a,b}$ so that, at equilibrium, the two oscillators have an average number of excitations $N_{a,b}=(e^{\hbar \omega_{a,b}/ k_B T_{a,b}}-1)
^{-1}$. The baths are assumed to be mutually independent. The system is then subjected to extra quantum noise, described by the input operators $\hat a^{\text{in}},\, \hat b
^{\text{in}}$ satisfying $\langle \hat a^{\text{in}, \dagger}(t)\hat a^{\text{in}}(t')\rangle=N_{a}\delta(t-t')$, and $\langle\hat a^{\text{in}}(t) \hat a^{\text{in}, \dagger}(t')\rangle=(N_{a}+1)\delta(t-t')$ 
(similar expressions hold for $\hat b^{\text{in}}$). Finally, the dissipation to the local baths occurs with rates $\kappa_{a,b}$. 
A broad range of systems fall into this description, ranging from quantum optics to solid-state devices, condensates and atomic ensembles. The resulting open dynamics can be 
described by the quantum Langevin equation $\dot{\hat u}=A \hat u(t)+\hat N(t)$ for the vector of the dimensionless quadrature operators 
$\hat u=(\hat q_a,\hat p_a,\hat q_b,\hat  p_b)^T$, where the drift matrix $A$ is given by
\begin{equation}\label{A}
A =
\begin{pmatrix}
-\kappa_a & \omega_a& 0 & 0 \\
-\omega_a& -\kappa_a& G & 0 \\
0 & 0& -\kappa_b & \omega_b \\
G & 0& -\omega_b & -\kappa_b \\
\end{pmatrix},
\end{equation}
while the noise vector is $\hat N=(\sqrt{2\kappa_a}\hat{q}_a^{\text{in}}, \sqrt{2\kappa_a}\hat{p}_a^{\text{in}}\sqrt{2\kappa_b}\hat{q}_b^{\text{in}},\sqrt{2\kappa_b}\hat{p}_b^{\text{in}})
^\mathrm{T}$. The linear character of the dynamics, together with the choice of initial Gaussian states, implies that the probability distribution describing the two oscillators is 
Gaussian at any time~\cite{GaussRev, MatteoRev}. Therefore a complete description of the system is given in terms of the second statistical moments of the quadrature operators, 
which can be arranged in the covariance matrix $\sigma$ of entries $\sigma_{ij}:=\langle\{\hat u_i(t),\hat u_j(t)\}\rangle/2-\langle\hat u_i(t)\rangle\langle\hat u_j(t)\rangle$. The first moments can straightforwardly be made null by a suitable displacement in the phase space and we assume, from now on, that this is indeed the case. The equation of motion for the covariance matrix reads 
\begin{equation}\label{SigmaEqMotion}
{\dot\sigma}=A \sigma + \sigma A^T+D\, , 
\end{equation}
where the diffusion matrix is given by $D=(1+2 N_a)\kappa_a\mathds{1}_a\oplus (1+2 N_b)\kappa_b\mathds{1}_b$. 
\par
The competition between the two baths at different temperatures causes the breaking of the detailed balance and brings
the system out of equilibrium~\cite{Esposito2}. Moreover, the unitary evolution generated by Eq.~\eqref{Hamiltonian} does not commute with the dissipative part of the dynamics. Therefore, the coherent coupling between the quantum oscillators affects the overall irreversibility. We assume the system to be always stable, in such a way that a unique non-equilibrium steady state, described by the stationary covariance matrix $\sigma_{\text s}$ such that 
$A \sigma_{\text s} + \sigma_{\text s} A^T = - D$, is eventually attained.

The open dynamics can be described in terms of Fokker-Plank equations for the Wigner function of the joint system and, provided that the symmetry of the variables under 
time-reversal is explicitly taken into account~\cite{Tome, Landi, SpinneyFord}, san analytical expression for $\Pi_\text{s}$ can be derived starting from Eq.~\eqref{RateEq}. Explicit calculations shown in Appendix B lead to the following simple expression for the stationary rate of entropy production
\begin{equation}\label{EntropyNESS}
\Pi_\text{s}= 2 \kappa_a \left( \frac{\langle\hat{q}_a^{2}\rangle_\text{s}+\langle \hat{p}_a^{2}\rangle_\text{s}}{2N_{a} +1}-1\right)   + 
2 \kappa_b \left( \frac{\langle\hat{q}_b^{2}\rangle_\text{s}+\langle \hat{p}_b^{2}\rangle_\text{s}}{2N_{b} +1}-1\right) \,, 
\end{equation}
where $ \langle\, \cdot \,\rangle_\text{s}$ specifies that the expectation values are taken at the stationary state. Since the first (second) term depends only on quantities labeled by 
$a$ ($b$) we dub it contribution $a$ ($b$) to the entropy production rate and call it $\mu_a$ ($\mu_b$). 
We thus have
\begin{equation}\label{Mu_ab}
\mu_k=2 \kappa_k \left( \frac{N_{k,\text{s}}+1/2}{N_k+1/2}-1\right),~~~(k=a,b)
\end{equation}
where we set $N_{a,\text{s}}=\langle\hat{a}^{\dagger}\hat{a}\rangle_\text{s}$ and $N_a=\langle\hat{a}^{\dagger}\hat{a}\rangle_\text{eq}$, and similarly for 
$\mu_b$. The main feature of Eq.~\eqref{Mu_ab} is that it links the irreversibility generated by the stationary process to the change in the amount of excitations 
carried by each oscillator with respect to the equilibrium value, thus expressing  production of entropy in very simple terms. 

\begin{figure}[t] 
{\bf (a)}~\hskip4cm{\bf (b)}
\includegraphics[scale=.33]{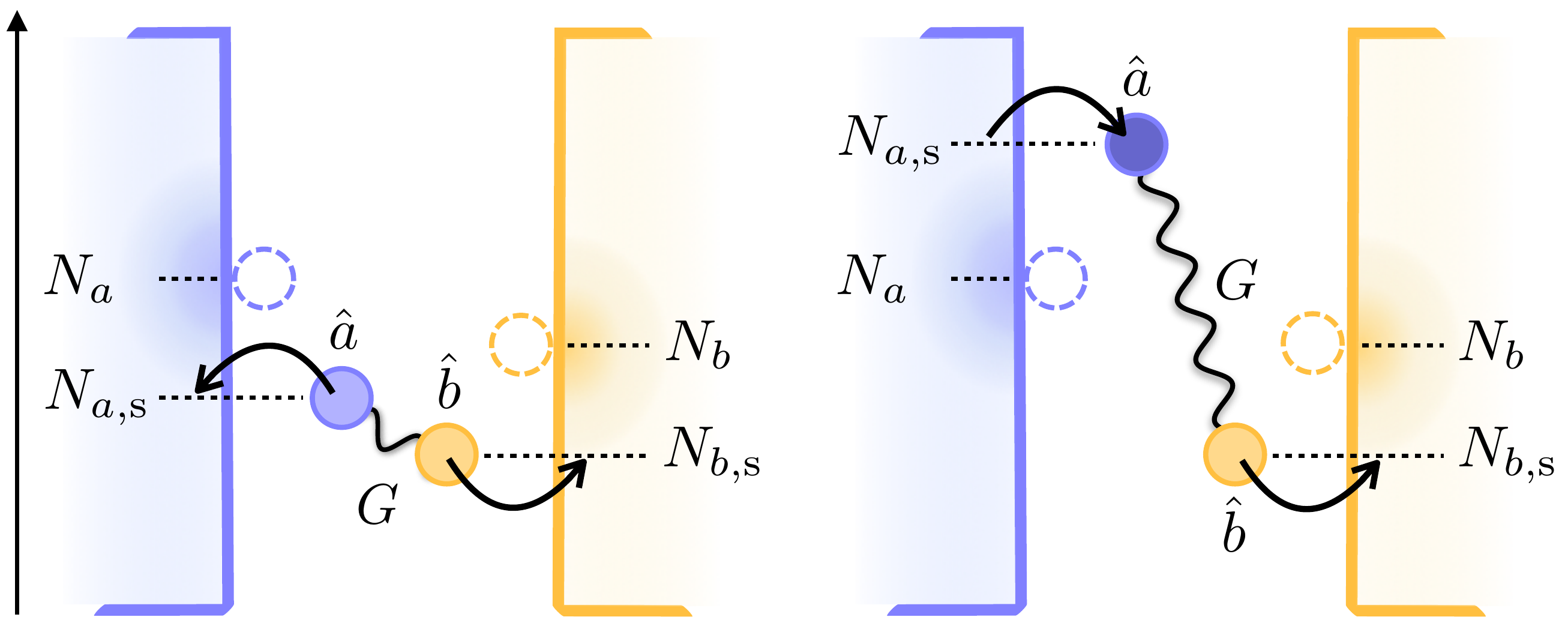}
\caption{The oscillators corresponding to modes $\hat a$ (blue) and $\hat b$ (yellow) are initially in thermal equilibrium with a number of excitations $N_a$ and $N_b$, 
respectively (dashed circles). By switching on the coupling $G$ they reach a stationary state characterized by occupations $N_{a,{\text s}}$ and $N_{b,{\text s}}$ (full circles). 
{\bf (a)}: Example of a forbidden stationary process where both occupations decrease with respect their equilibrium values, thus leading to $\Pi_{\text s}<0$. {\bf (b)}: Entropy can 
still locally decrease ($\mu_b<0$) as a consequence of a reduction in the excitations $N_{b,{\text s}}<N_b$, but this necessitates excitations to be accumulated in mode $\hat a$ 
(darker blue circle). 
\label{f:Sketch}}
\end{figure}

If the system is noninteracting, each oscillator equilibrates with its own bath and from Eq.~\eqref{Mu_ab} we see that $\Pi_{\text{s}}$ 
identically vanishes. Second, as $\Pi_\text{s}=\mu_a+\mu_b\ge0$, from Eq.~\eqref{Mu_ab} we conclude that no process leading 
at the same time to $N_{a,\text{s}} < N_a$ and $N_{b,\text{s}} < N_b$ can occur: the thermodynamic arrow of time is translated in a constraint on the final occupations of the two oscillators. 
An instance of forbidden process is sketched in Fig.~\ref{f:Sketch} {\bf (a)}. However, nothing prevents a local 
reduction of entropy, e.g. $\mu_b<0$ as shown in panel {\bf (b)}, as long as it is (over)compensated by an increase of the other contribution 
$\mu_a>-\mu_b$. Such condition entails $N_{b,{\text s}}<N_b$ and thus corresponds to the cooling one oscillator assisted by the interaction. This also implies 
that, singularly taken, neither $\mu_a$ nor $\mu_b$ can be interpreted as an entropy production. 

\par 
Looking at Eq.~\eqref{EntropyNESS} we notice that there is no explicit dependence of $\Pi_{\text s}$ on the off-diagonal elements of the covariance matrix. Correlations 
between the two modes are hidden in the full expression of the expectation values. It would be desirable to have an alternative form for $\mu_{a,b}$, 
where the role of the correlations established at the steady state is made explicit. Such an expression can actually be derived (calculations are reported in Appendix C)
and is given by
\begin{equation}\label{Mu_nondiag} 
\mu_a= \frac{G}{N_a+1/2}\langle \hat{p}_a \hat{q}_b\rangle_{\text s},~\mu_b= \frac{G}{N_b+1/2}\langle \hat{q}_a \hat{p}_b\rangle_{\text s},
\end{equation}
where the we have $\langle \hat{p}_a \hat{q}_b\rangle_{\text s}=[\sigma_{\text s}]_{23}$ and $\langle \hat{q}_a \hat{p}_b\rangle_{\text s}=[\sigma_{\text s}]_{14}$. 
From Eq.~\eqref{Mu_nondiag}, we explicitly see that $\Pi_{\text s}$ vanishes for uncoupled systems, since each oscillator independently equilibrates with its 
own bath. Eq.~\eqref{Mu_nondiag} links in a quantitative way the irreversibility of the transformation with some correlation function of the dynamical variables. The link 
between the entropy production and the correlations shared by the oscillators will be further explored in Sec.~\ref{s:EntCorr}, where the amount of total and quantum 
correlations is quantified.

\section{analysis of the stationary entropy production rate}\label{s:EntProd}

In this Section we give a full account of the  behavior of the stationary entropy production. 
For the sake of convenience, all the frequencies have been rescaled by 
$\omega_b$, so that we deal with dimensionless quantities. 
However, in order to avoid redundancies, the rescaling will be omitted and the same notation kept, except from the figures and the related captions, where the relevant 
quantities are explicitly shown in units  of $\omega_b$. 
\begin{figure}[t] 
\includegraphics[scale=.49]{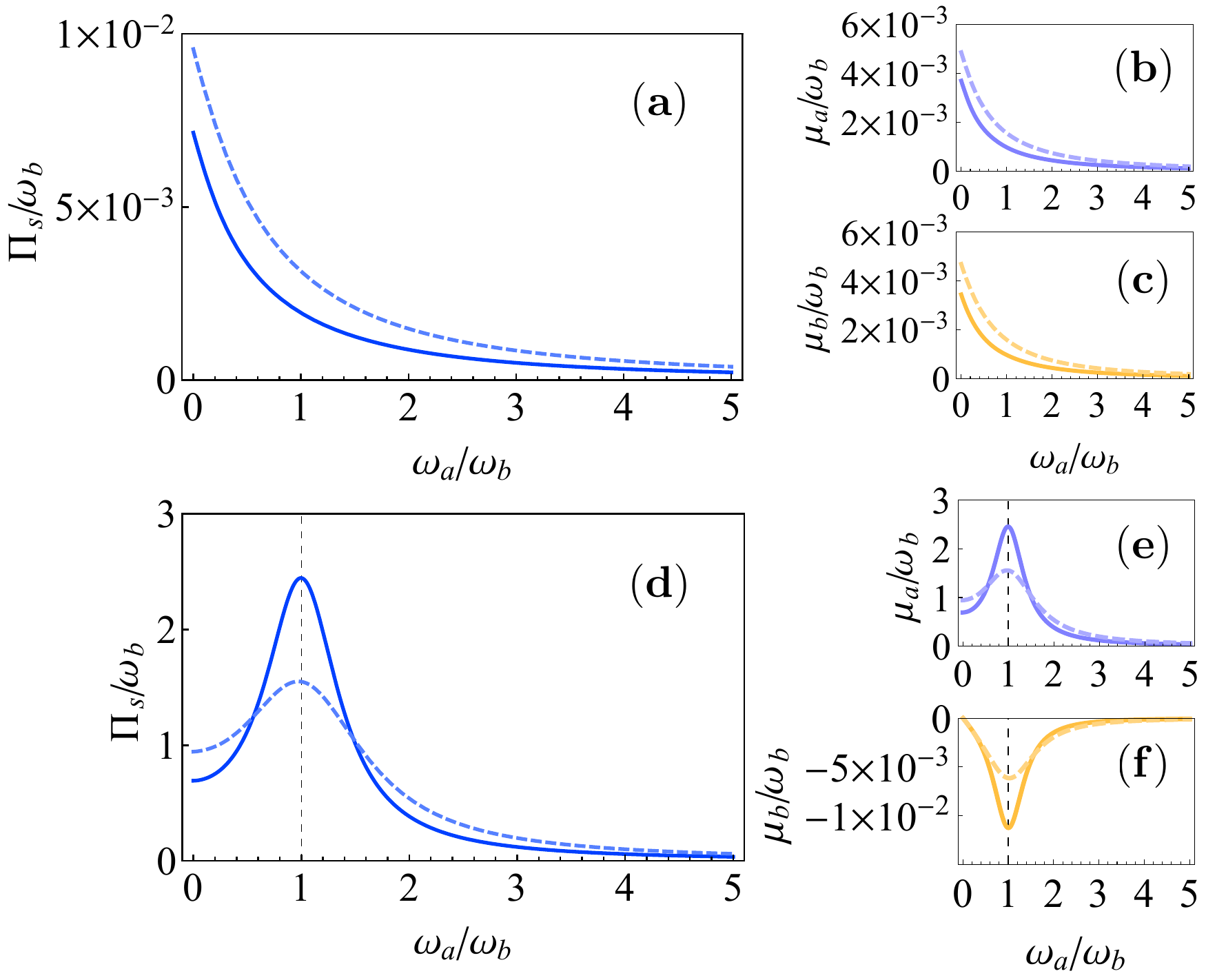}
\caption{Entropy production rate $\Pi_{\text s}/\omega_b$ \textbf{(a)} and its two contributions $\mu_a/\omega_b$ \textbf{(b)} and $\mu_b/\omega_b$ \textbf{(c)} 
against the ratio of the two frequencies; solid curves correspond to $\kappa_a=\kappa_b=0.2\omega_b$, while dashed ones to $\kappa_a=0.2\omega_b$ and $\kappa_b=0.5\omega_b$. 
Other parameters are $N_a=N_b=0$ and $G=0.1\omega_b$. Panels \textbf{(d)}, \textbf{(e)}, \textbf{(f)} represent the same plots as \textbf{(a)}, \textbf{(b)}, \textbf{(c)} when introducing an
imbalance in thermal excitations $N_a=0$ and $N_b=100$.
\label{f:PlotPI1}}
\end{figure}
\par
\begin{figure}[b] 
\includegraphics[scale=.51]{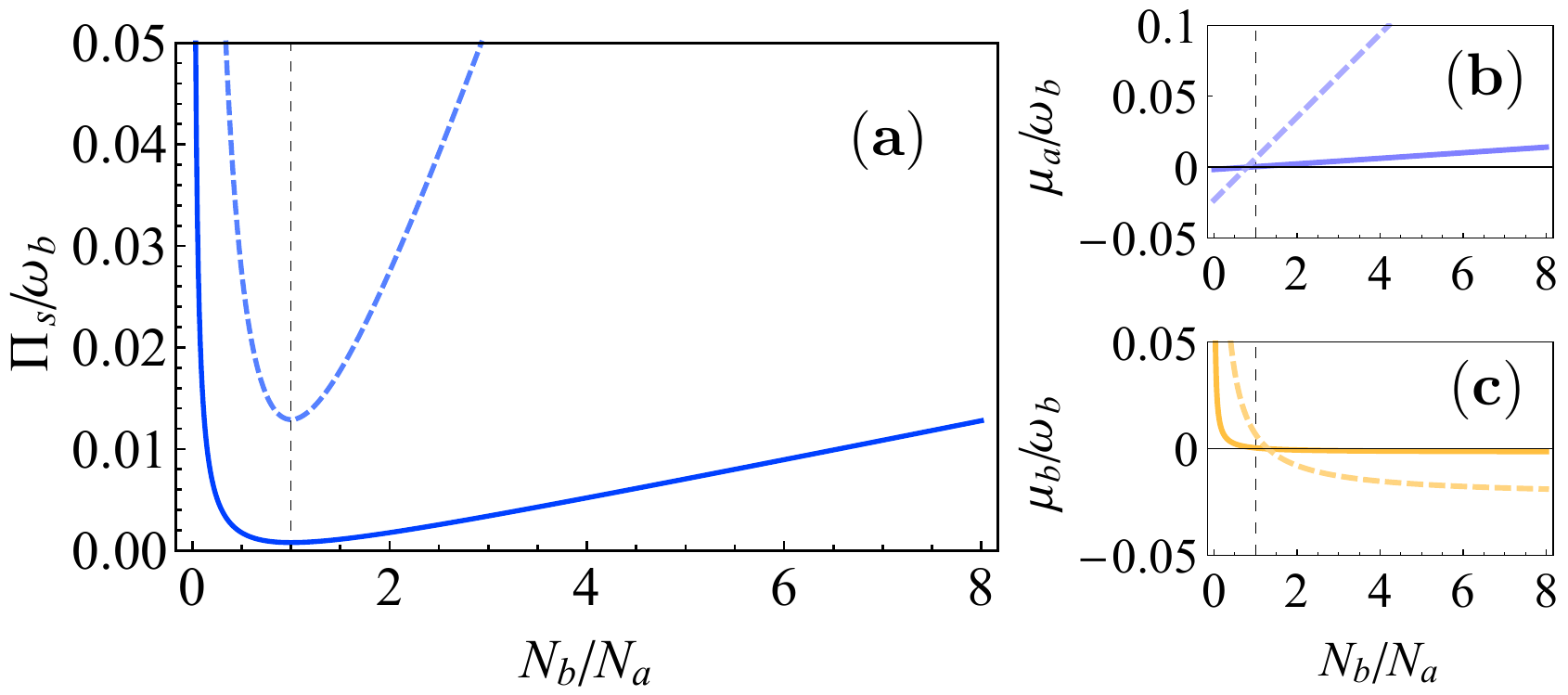}
\caption{Entropy production rate $\Pi_{\text s}/\omega_b$ \textbf{(a)} and its two contributions $\mu_a/\omega_b$ \textbf{(b)} and $\mu_b/\omega_b$ \textbf{(c)} 
against the ratio $N_b/N_a$.  The oscillators have the same frequency $\omega_a=\omega_b$, $\kappa_a=0.2\omega_b$ and $\kappa_b=0.5\omega_b$, and either coupled 
with strength $G=0.05\omega_b$ (solid curves) or $G=0.2\omega_b$ (dashed curves).
\label{f:PlotImbalance}}
\end{figure}

In Fig.~\ref{f:PlotPI1} we show the stationary entropy production rate $\Pi_\text{s}$, together with its components $\mu_{a,b}$, against the rescaled frequency $\omega_a$. In panels \textbf{(a)}-\textbf{(c)} the reservoirs are in the ground state ($N_a=N_b=0$) and we see that $\mu_a$ and $\mu_b$ are both positive and very 
similar (although not equal). This is because the steady-state occupations can only increase with respect to their initial value and, by looking at Eq.~\eqref{Mu_ab}, so must the entropy.
If we then consider some initial thermal occupation in one oscillator, as shown in Fig.~\ref{f:PlotPI1} \textbf{(d)}-\textbf{(f)} for the case $N_b>0$, we see that 
$\Pi_{\text s}\approx \mu_a$, featuring a distinctive peak at $\omega_a=1$. Correspondingly, $\mu_b$ displays a negative dip. The significant difference in magnitude between 
$\mu_a$ and $\mu_b$ ensures the overall positivity of $\Pi_{\text s}$. 
By comparing panels \textbf{(a)} and \textbf{(d)} we immediately notice that the introduction of some imbalance between the initial populations of the oscillators causes $\Pi_\text{s}$ to 
grow, thus witnessing irreversibility due to transport, since the coupled oscillators now mediate a net heat flux between the two baths. Moreover, in light of Eq.~\eqref{Mu_ab}, a steady negative value of $\mu_b$ implies a reduction of steady excitations $N_{b,\text{s}} < N_b$ and thus an effective cooling of oscillator $b$ (as sketched in Fig. \ref{f:Sketch}, right panel). 
The maximum (minimum) assumed by $\mu_a$ ($\mu_b$) at $\omega_a=1$, namely when the oscillators have identical frequencies, can be understood as follows. For $\omega_a\approx1$, and provided that $G<\omega_a$, we can move to the rotating frame and apply the rotating-wave approximation, so that the interaction Hamiltonian takes the form 
$H_{I}\propto \hat a^{\dagger} \hat b+\hat a\hat b^{\dagger}$. The latter is a pure exchange interaction and hence is optimal for heat transfer, thus explaining why the degree of irreversibility is large~\cite{Briegel}. These features will be discussed in more detail in Sec.~\ref{s:OptoCorr} when addressing the cavity-assisted cooling of a mechanical resonator.  
On the other hand, from Fig. \ref{f:PlotPI1} we see that $\Pi_{\text s}$ tends to zero for $\omega_a\gg1$. This is because when the  oscillators are far-off resonance they are effectively decoupled, so that each of them thermalizes to its own bath. 
Finally, we can also inspect the role played by the dissipation rates $\kappa_{a,b}$ on the irreversible entropy production. The solid curves in the plots correspond to identical loss rates 
$\kappa_a=\kappa_b$ while the dashed curves refer to the case of different values ($\kappa_a=0.2,\,\kappa_b=0.5$), and we can see that a general feature is the broadening of $\Pi_{\text s}$ with the losses.
When the temperature $T_{a,b}$ of the baths and the frequencies $\omega_{a,b}$ are such that the initial number of thermal excitations is the same, i.e. $N_a=N_b=N$, \textit{the entropy production rate turns out to be independent on N}, which is a clear feature of the enforced symmetry between the two subsystems. Also, from Fig. \ref{f:PlotImbalance} 
we see that $\Pi_\text{s}$ achieves its minimum for $N_a=N_b$, while when $N_b$ exceeds $N_a$,  $\Pi_\text{s}$ grows linearly with respect to $N_b$, i.e. proportionally to the imbalance 
in populations. For $N_b/N_a<1$ we see a an abrupt increase in the entropy production rate, which however remains finite for $N_b=0$. 

Some more quantitative insight can be gathered in the small coupling limit, after expanding $\mu_{a,b}$ in power series of $G$, getting the following expressions
\begin{widetext}
\begin{align}\label{ExpG}
\mu_a&=\frac{ G^2 \kappa_{\text{tot}} \left[1+\kappa_{\text{tot}}^2+\omega_a^2 -2\omega_a(2N_a+1)+2N_b\left(1+\kappa_{\text{tot}}^2+\omega_a^2\right)\right]}{(2 N_a+1) 
\left[2 \omega_a^2 (\kappa_{\text{tot}}^2-1)+\left(\kappa_{\text{tot}}^2+1\right)^2+\omega_a^4\right]}+\mathcal{O}\left(G^4\right) \nonumber, \\
\\
\mu_b&=\frac{ G^2 \kappa_{\text{tot}} \left[1+\kappa_{\text{tot}}^2+\omega_a^2 -2\omega_a(2N_b+1)+2N_a\left(1+\kappa_{\text{tot}}^2+\omega_a^2\right)\right]}{(2 N_b+1) 
\left[2 \omega_a^2 (\kappa_{\text{tot}}^2-1)+\left(\kappa_{\text{tot}}^2+1\right)^2+\omega_a^4\right]}+\mathcal{O}\left(G^4\right) \nonumber,
\end{align}
\end{widetext}
where we set $\kappa_{\text{tot}}=\kappa_a +\kappa_b$. We can see that the first non vanishing term is quadratic in $G$, and in addition it can be checked that $\mu_{k}~(k=a,b)$ is an even function of $G$. We also notice that in such a  small coupling regime, $\mu_a$ and $\mu_b$ are simply \textit{related to each other by the swap of 
the thermal excitations} $N_{a,b}$: the imbalance between $N_a$ and $N_b$ has a clear effect on the two components $\mu_{a,b}$ and dictates which of the oscillators will reduce its local entropy and which one will compensate. The coherent coupling between the two oscillators alters this simple picture, since already by including the fourth-order 
term this ceases to be true. If we then expand $\mu_{a,b}$ for large values of $\omega_a$, we obtain the following expressions for the asymptotic behavior of the tails 
\begin{equation}\label{ExpOmega}
\begin{aligned}
\mu_a &= \frac{1}{\omega_a^2}\frac{G^2\kappa_{\text{tot}}(1+2N_b)}{1+2N_a} + \mathcal{O}\left(\frac{1}{\omega_a^3}\right),\\
\mu_b &= \frac{1}{\omega_a^2}\left[\frac{G^2\kappa_{\text{tot}}(1+2N_a)}{1+2N_b} + \frac{G^4 \kappa_b}{2(\kappa_b^2+1)}\right]+ \mathcal{O}\left(\frac{1}{\omega_a^3}\right).
\end{aligned}
\end{equation}
For large values of $\omega_a$, the entropy production rate (at the leading order) decays as $\omega_a^{-2}$. In the small-coupling limit, when the term 
proportional to $G^4$ in Eq.~\eqref{ExpOmega} can be neglected, $\mu_{a,b}$ can be mapped into each other by swapping the thermal excitations, as previously argued. Finally, we 
notice that both contributions in Eq.~\eqref{ExpOmega} are strictly positive, which may seem in contradiction to what shown in Fig.~\ref{f:PlotPI1} \textbf{(f)}, where $\mu_b$ takes negative
values. However, if we retained the next-order term in the expansion, we would find that $\mu_b$ approaches zero from below for $\omega_a \gg 1$, in agreement with the plot.  
\begin{figure}[h!] 
\includegraphics[scale=.49]{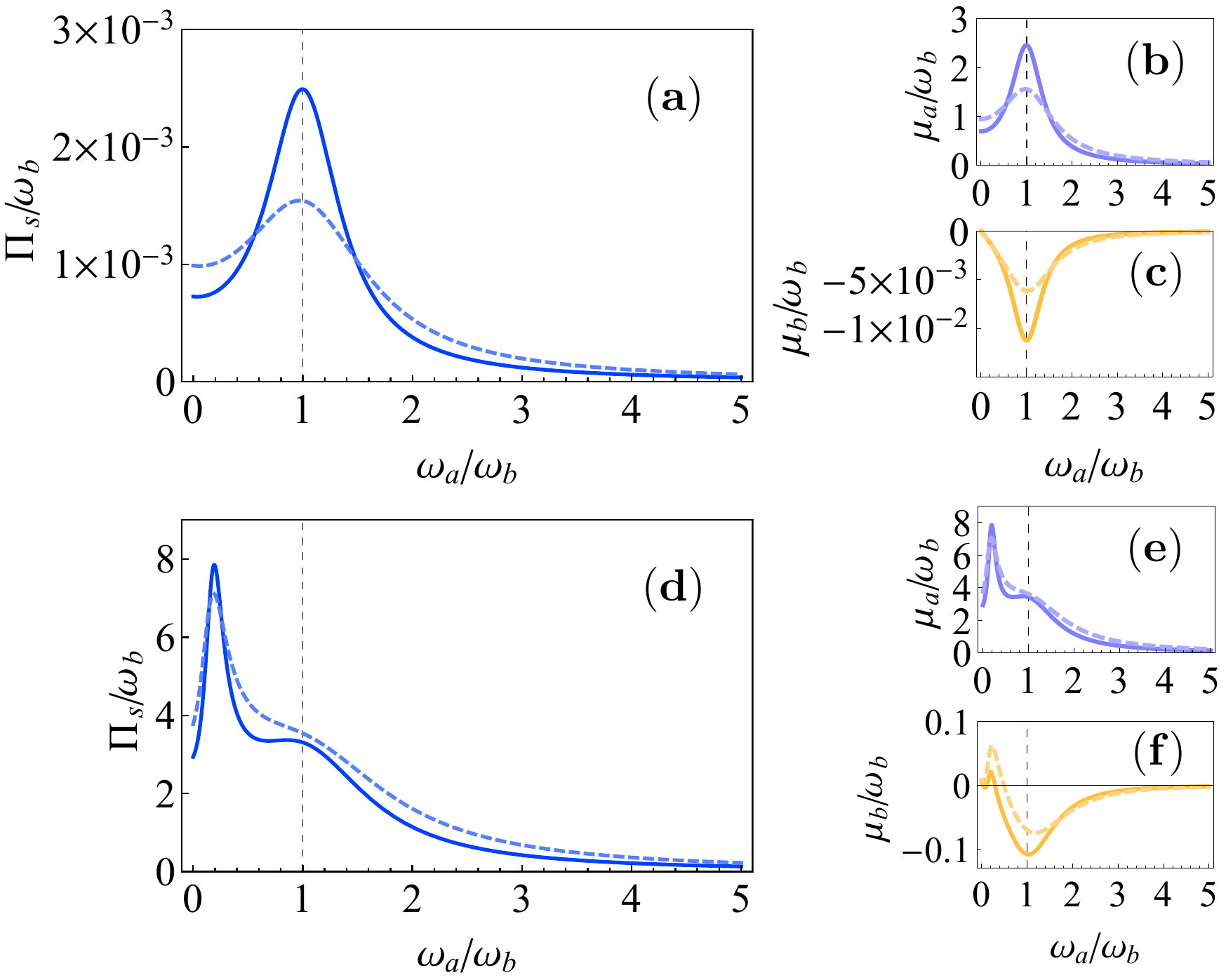}
\caption{We plot $\Pi_{\text s}/\omega_b$ [panel \textbf{(a)}], $\mu_a/\omega_b$ [panel \textbf{(b)}] and $\mu_b/\omega_b$ [panel \textbf{(c)}] against the ratio of the two frequencies
for different values of the coupling; solid lines correspond to $\kappa_{a,b}=0.2\omega_b$, while dashed ones to $\kappa_a=0.2\omega_b$ and $\kappa_b=0.5\omega_b$. 
Other parameters are $N_a=0$, $N_b=10$ and $G=0.01\omega_b$. Panels \textbf{(d)}-\textbf{(f)} show the same plots as panels \textbf{(a)}-\textbf{(c)} but for $G=0.6\omega_b$.
\label{f:PlotPI2}}
\end{figure}
\par
We conclude the description of the features of $\Pi_{\text s}$ by inspecting its dependence on the coupling strength $G$. By comparing Fig.~\ref{f:PlotPI2} \textbf{(a)}, corresponding to $G=0.01$, with  Fig.~\ref{f:PlotPI1} \textbf{(d)}, which is  for $G=0.1$, we further confirm the quadratic scaling in Eq.~\eqref{ExpG}, since $\Pi_{\text s}$ retains the 
very same shape but drops by two orders of magnitude. On the other hand, in Fig. \ref{f:PlotPI2} \textbf{(d)} we show an instance of the strong coupling regime, where the strength of the 
coherent coupling exceeds the dissipation rates and is comparable with the oscillation frequencies. Now, $\Pi_{\text s}$ exhibits a second peak, which is a clear signature of hybridization.

\subsection{Identical oscillators and identical baths}
If the baths have the same temperature $T$ and the oscillators the same frequency $\omega$, which results in the same number of excitations $N=(e^{\hbar \omega/ k_B T}-1)^{-1}$, we can directly relate the entropy production rate to a single element of the covariance matrix of the oscillators' state. We refer to the entropy production rate 
in this case as $\tilde \Pi_{\text s}$, whose expression is given by  $tilde \Pi_{\text s}=\frac{G}{\omega} \frac{\kappa_{\text{tot}}}{N+1/2}\langle \hat{q}_a \hat{q}_b\rangle_{\text s}$, which shows an explicit proportionality to the matrix element $\langle \hat{q}_a \hat{q}_b\rangle_{\text s}=[\sigma_{\text s}]_{13}$. By using the explicit form of such covariance matrix entry, we have
\begin{align}\label{PI_K1K2}
\tilde \Pi_{\text s}
&=\frac{G^2 \kappa_{\text{tot}} \left[G^2 \left(\kappa_{\text{tot}}^2-3\kappa_a\kappa_b+1\right)+4 \kappa_a \kappa_b\chi_{ab}\right]}{2 \left(\chi_{ab}-G^2\right) 
\left[G^2+\kappa_a \kappa_b \left(\kappa_{\text{tot}}^2+4\right)\right]},
\end{align}
where we have set $\chi_{ab}=\left(\kappa_a^2+1\right)\left(\kappa_b^2+1\right)$. Eq.~\eqref{PI_K1K2} shows that, In agreement with what previously stated, for two identical oscillators in the absence of any thermal gradient, the entropy production is independ of $N$. Furthermore, if we impose $\kappa_a=\kappa_b=\kappa$ (and dub the corresponding entropy production $\tilde \Pi_{\kappa}$) we obtain the simple expression 
\begin{equation}\label{PI_Kappa}
\tilde \Pi_{\kappa}=\frac{G^2 \kappa  (\kappa ^2+1)}{(\kappa ^2+1)^2-G^2}.
\end{equation}
Having two identical oscillators dissipating at the same rate into identical baths, the open system is in its most symmetric configuration, and the resulting stationary state is the closest 
possible to equilibrium. Accordingly, one can verify that in this scenario the entropy production is minimised. Moreover, a detailed calculation shows that, in this case, the  oscillators contribute equally
to the production of entropy, i.e., $\tilde \mu_a=\tilde \mu_b=\tilde \Pi_{\kappa}/2$.

\section{Entropy production rate, total and quantum correlations}\label{s:EntCorr}
\subsection{Quantifying correlation via the R\'enyi-2 entropy}
The mutual information $\mathcal{I}(\varrho_{a:b})=S(\varrho_{a})+S(\varrho_{b})-S(\varrho_{ab})$ quantifies the total amount of correlations shared by two systems, the 
rationale being that only when two systems are completely uncorrelated their joint entropy equals the sum of the reduced entropies~\cite{VlatkoRevEntropy}. While the standard way to quantify entropy is to use the von Neumann entropy, the Gaussian nature of  the states at hand here, which are completely characterized by the two-mode covariance matrix
\begin{equation}\label{sigma}
\sigma_{ab}=
\begin{pmatrix}
\sigma_a \, & c_{ab} \\
c_{ab}^T &  \sigma_b
\end{pmatrix},
\end{equation} 
legitimates the use of another expression for the entropy. This is given in terms of the Shannon entropy of the Wigner distribution 
$\mathcal{W}(\sigma_{ab})=e^{-u^T\sigma_{ab}^{-1}u/2}/(\pi^2\sqrt{\text{det}\sigma_{ab}})$, and reads  
\begin{equation}
S(\sigma_{ab})=\int_{\mathbb{R}^4}\mathrm{d}^4 u\mathcal{W}(\sigma_{ab})\log\mathcal{W}
(\sigma_{ab}),
\end{equation}
where $u=(q_a, p_a, q_b,  p_b)^T$ is the vector of phase-space variables. For Gaussian distributions the expression can be easily evaluated and gives  
\begin{equation}\label{ShanWig}
S(\sigma_{ab})=\frac12 \log (\det \sigma_{ab}).
\end{equation}
Ref.~\cite{Adesso} pointed out that $S(\sigma_{ab})$ coincides (up to an additive constant) with the generalized R\'enyi 
entropy of order 2, $S_2(\varrho)=-\log \tr{\varrho^2}$, and proved that it enjoys all the properties required to be a legitimate entropy measure, 
such as strong subadditivity. Such a correspondence is discussed in detail in Appendix A.
\par
We can then define the mutual information as $\mathcal{I}(\sigma_{a:b})=S(\sigma_{ab}\vert\vert\sigma_a \oplus \sigma_b)$, namely as
the Kullbac-Leibler divergence between the Wigner function of the joint state and the product of those associated with the reduced ones. From Eq. (\ref{ShanWig}) it immediately follows that
\begin{equation}\label{IG}
\I(\sigma_{a:b})=\frac12 \log \left(\frac{\det\sigma_{ab}}
{\det\sigma_{a}\det\sigma_{b}} \right) \,.
\end{equation}
Finally, also the amount of quantum correlations between the two modes can be expressed in terms of the R\'enyi-2 entropy. The quantum discord (with respect 
to measurements made  on mode $\hat b$) is  defined as the difference
\begin{equation}\label{DG}
\D(\sigma_{a\vert b})=\I(\sigma_{a:b})-\mathcal{J}(\sigma_{a\vert b}),
\end{equation}
between the mutual information $\I(\sigma_{a:b})$ and the one-way classical correlations $\mathcal{J}(\sigma_{a\vert b})=\sup_{\pi_b(X)}\{S(\sigma_a)-
\int\text{d}X p_X S(\sigma_{a\vert X}^{\pi_b}) \}$, where the maximization is taken over all the possible measurements $\hat\pi_b(X)\ge0$ such that 
$\int \text{d}X\hat\pi_b(X)=\mathds{1}_b$, implemented on $\hat b$. If we restrict our attention to Gaussian measurements, we can rewrite Eq.~\eqref{DG} as 
$\D(\sigma_{a\vert b})= S(\sigma_b)-S(\sigma_{ab})+\inf_{\pi_b}S( \sigma_a^{\pi_b})$, and the minimization can be performed analytically. All the details 
are reported in Appendix D, together with the fully-fledge expression of $\D(\sigma_{a\vert b})$. In what follows we will employ such figures of merit to 
characterize the correlations between the two modes $\hat a$ and $\hat b$ at the stationary state. When no further specifications are needed, we will refer 
to the R\'enyi-2 mutual information and discord simply as $\I$ and $\D$, respectively. 

\subsection{Relationship between entropy production rate and mutual information}
\begin{figure}[t] 
\includegraphics[scale=.51]{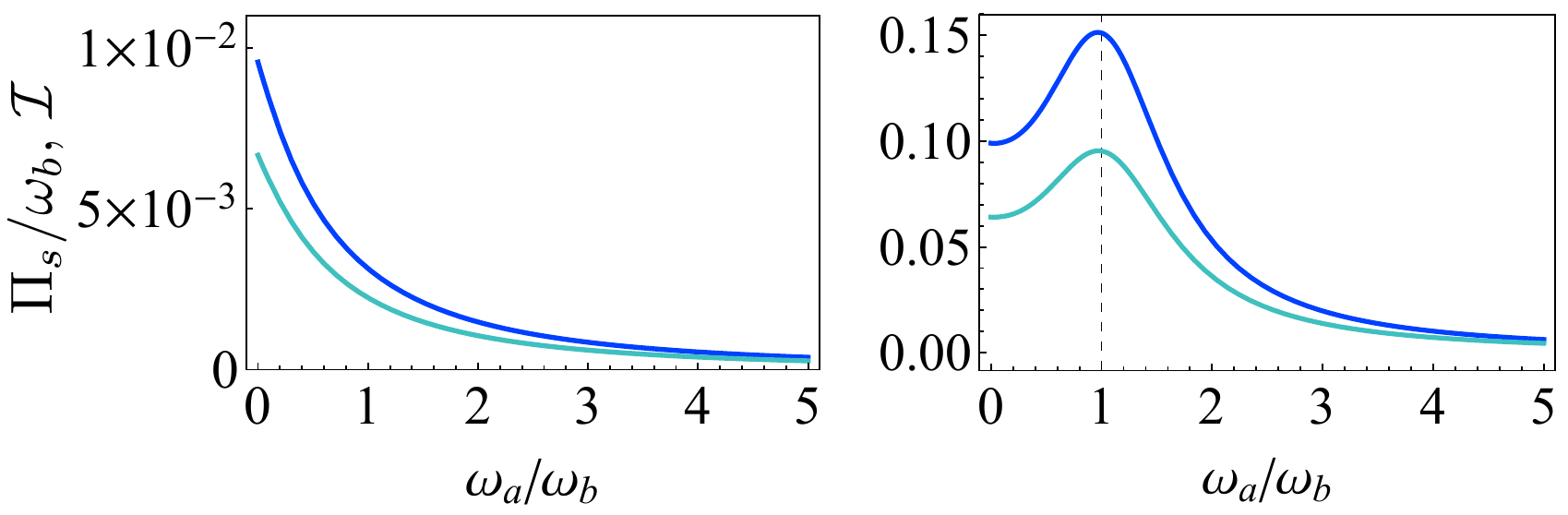}
\caption{Comparison between the rescaled entropy production rate $\Pi_{\text s}/\omega_b$ (blue), the mutual information $\I$ (green) and the quantum discord
$\D$ (magenta) as a function of the ratio of the two frequencies $\omega_a/\omega_b$, for $G=0.1\omega_b$, $\kappa_a=0.2\omega_b$, $\kappa_b=0.5\omega_b$, 
$N_a=0$ and either $N_b=0$ (left) or $N_b=10$ (right). 
\label{f:PlotCorr1}}
\end{figure}
The full expression for the mutual information between the two oscillators at the steady state is extremely involved, and it will not be reported here. However, we must 
stress that in general its form bears no resemblance with the entropy production rate: $\Pi_{\text s}$ is a rational function, and in particular the ratio between two fourth order 
polynomials in $G$, while from Eq.~\eqref{IG} we clearly see that $\I$ is logarithmic. Furthermore, at variance with $\Pi_{\text s}$, the mutual information cannot 
be reduced to the sum of two separate contributions, each ascribable to a single oscillator. These facts reflect the different origin of the two terms. On the other hand, 
thanks to Eq.~\eqref{Mu_nondiag} we know that $\Pi_{\text s}$ is explicitly linked to position-momentum correlations among the oscillators, so that one may wonder about 
a possible connection between the rate of entropy produced and the correlations present within the system.  
In fact, by comparing $\Pi_{\text s}$ and $\I$, as done in Fig. \ref{f:PlotCorr1}, we notice a striking similarity: despite their different functional form, $\I$ is clearly a one-to-one function of 
$\Pi_{\text s}$. This represents a first evidence that the irreversibility generated by the stationary process and the total amount of correlations shared between the two modes
are tightly related.
\par
\begin{figure}[h!] 
\includegraphics[scale=.52]{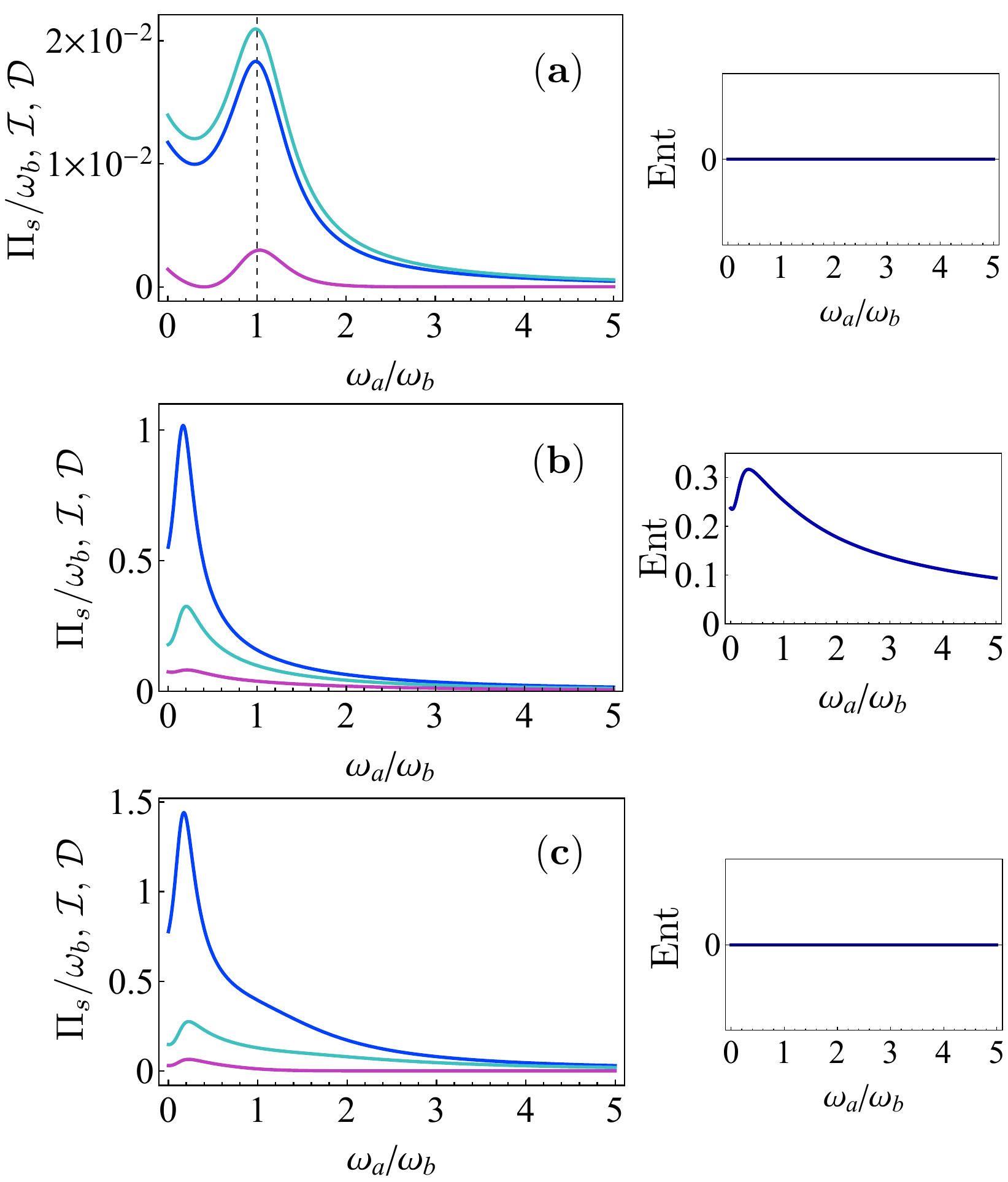}
\caption{Comparison between the steady-state (rescaled) entropy production rate $\Pi_{\text s}$ (blue), the mutual information $\I$ (green water), and the quantum discord 
$\D$ (magenta) as a function of the ratio of the two frequencies $\omega_a/\omega_b$ (left column) together with the entanglement between the mechanical and optical 
field (right column) quantified by the logarithmic negativity. \textbf{(a)} is for $G=0.1\omega_b$, $\kappa_a=\kappa_b=0.2\omega_b$ and $N_a=0$ and $N_b=1$; in \textbf{(b)} 
$g=0.6\omega_m$, $\kappa_a=0.2\omega_b$, $\kappa_b=0.5\omega_b$, $N_a=N_b=0$. Finally in \textbf{(c)} same parameters as \textbf{(b)} except for $N_b=1$.
\label{f:PlotCorr2}}
\end{figure}

  
When the oscillators are uncoupled or far-off-resonance, and thus effectively decoupled, they separately reach thermal equilibrium and the total state of the system consists 
of locally thermal states, yielding vanishing values of both  $\Pi_{\text s}$ and $\I$. Another remarkable feature shared by the two figures of merit is that, 
when the number of thermal excitations $N_a=N_b=N$ is the same, both $\I$ and $\Pi_{\text s}$ turn out to be independent on $N$. 
As one can imagine, this is however not the case for the amount of entanglement, if any, shared by the two oscillators, which crucially depends on the thermal
occupations and is rapidly depleted when increasing $N_{a,b}$ (even for $N_a=N_b=N$). In Fig. \ref{f:PlotCorr2} \textbf{(b)} we computed the logarithmic negativity 
when both oscillators dissipate into the vacuum ($N=0$); if we raised the thermal excitations to $N=1$ we would find $\Pi_{\text s}$ and $\I$ unchanged but the
entanglement completely spoiled. Actually, as shown in panel \textbf{(c)}, already lifting one of the two thermal occupation numbers ($N_b=1$) causes a complete loss of 
the entanglement (notice however that $\Pi_{\text s}$ increases thanks to the introduction of a thermal gradient).
Furthermore, as we can appreciate from Fig. \ref{f:PlotCorr2} \textbf{(b)} and \textbf{(c)}, by increasing the coupling strength $G$ the agreement between the two figures of 
merit gets worse. For big imbalances in the initial thermal  populations and strong couplings  $\mathcal{I}$ and $\Pi_{\text s}$ can develop different features and in some 
instance also the qualitative agreement may be lost.
\par
In order to gain a more quantitative understanding about the connection highlighted in Fig. \ref{f:PlotCorr1} we can expand both $\Pi_{\text s}$ and $\mathcal{I}$ in power 
series of $G$ [for $\Pi_{\text s}$ this amount to take the sum of the expansions of $\mu_{a,b}$ in Eq.~\eqref{ExpG}], obtaining the following simple relation
\begin{equation}\label{MainResult}
\I=\frac{\Pi_{\text s}}{2 \kappa_{\text {tot}}} + \mathcal{O}(G^4) \, .
\end{equation}
Eq.~\eqref{MainResult} is one of our main results, since it establishes a {\it proportionality between the total amount of correlations and the irreversibility of the dissipative 
process}. Provided that the coupling strength is not too big with respect to the frequency $\omega_b$, $\I$ and $\Pi_{\text s}$ turn out to be two sides of the same coin: 
more correlations in the system come at the price of an increased irreversibility and, conversely, the amount of entropy generated by the dissipative process is proportional 
to the correlations established by it. We must stress that this expansion does not necessarily entails a classical-like limit, and also systems displaying pronounced quantum features, 
e.g. stationary entangled states of the two oscillators, fulfill  Eq.~\eqref{MainResult}, as further discussed in the next Section.
\par
Finally, for the particular case of two identical oscillators dissipating with the same rate $\kappa$ into baths at the same temperature $T$, the mutual information $\I$ takes 
the form
\begin{equation}\label{MI_Kappa}
\tilde{\mathcal{I}}_{\kappa}=\frac{1}{2} \log \left[\frac{4 \left(G^2 K_- +2 K_+^2\right)^2}
{\left(K_+^2-G^2\right) \left(G^4+8 G^2 K_- +16 K_+^2\right)}\right] \, ,
\end{equation}
where we set $K_{\pm}=\kappa ^2\pm1$. The corresponding expression of the entropy production rate has been given in Eq.~\eqref{PI_Kappa}. Eq.~\eqref{MI_Kappa} 
is an exact result, valid for any value of the coupling strength, independent on $N$. By comparing Eqs.~\eqref{PI_Kappa} and Eq.~\eqref{MI_Kappa} we can conclude 
that, even for the specific case examined, their full expressions bear little resemblance.

\subsection{Relationship between entropy production rate and quantum discord}
We can also investigate the relation between $\Pi_{\text s}$ and the amount of quantum correlations, as quantified by the R\'enyi-2 discord $\D$, shared by the 
two oscillators at the steady state. From Fig. \ref{f:PlotCorr2} we see how the discord, which corresponds to the magenta curve, although assuming lower values, 
shares some common feature with $\Pi_{\text s}$. In particular from panel \textbf{(a)} we notice a pronounced similarity. We can expand $\D$ in series of the
coupling strength (given the complexity of the expression we consider same initial thermal occupation $N_a=N_b=N$ and same rates $\kappa$) to get the
following expression 
\begin{equation}\label{Discord}
\D=\frac{\Pi_{\text s}}{4\kappa_{\text{tot}}(N+1)}+\mathcal{O}(G^4) \, ,
\end{equation}
where $\kappa_{\text{tot}}=2\kappa$, and for completeness we also report the corresponding expression of  $\Pi_{\text s}$ 
\begin{equation}
\Pi_{\text s}=\frac{2 \kappa_{\text{tot}} G^2}{ \kappa_{\text{tot}}^2+(\omega_a+1)^2}\, .
\end{equation}
For small enough values of $G$ {\it the rate of irreversibility in the dynamics is  proportional to the quantum correlations between the two modes}. We point out 
that, at variance with $\Pi_{\text s}$ and $\I$, the discord does depend on the initial number of thermal excitations. Moreover, a comparison between Eq.~\eqref{MainResult}
and Eq.~\eqref{Discord} shows that, when both the baths are in the ground state, we have $\D=\I/2$.
\begin{figure}[h!] 
\includegraphics[scale=.62]{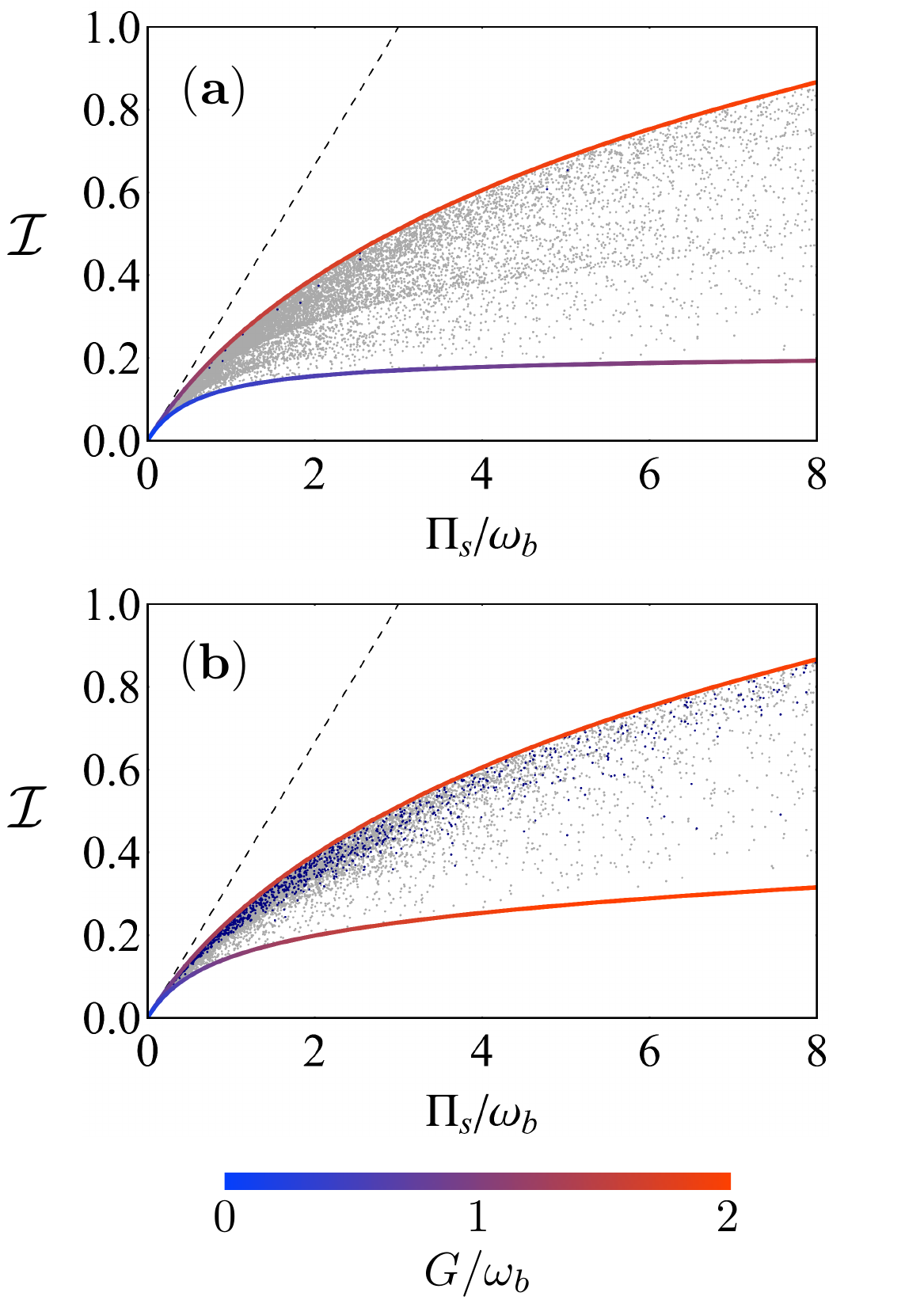}
\caption{Mutual information $\I$ against the rescaled irreversible entropy production rate $\Pi_{\text s}/\omega_b$ for randomly generated states. Each point corresponds 
to a state obtained by a uniform sampling from $\omega_a\in [0,3\omega_b]$, $G\in [0,2\omega_b]$, and either $N_{a,b}\in [0,10]$ \textbf{(a)} or $N_{a,b}\in [0,1]$ \textbf{(b)}.
Dissipation rates are kept fixed at $\kappa_a=0.5\omega_b$, $\kappa_b=\omega_b$. Points corresponding to separable states are shown in gray, while entangled 
states are marked in blue. The upper and lower bounds are parametric curves obtained by varying the coupling $G$ from $G_{\text{min}}=0$ (blue) to a maximum value 
$G_{\text{max}}=2\omega_b$ (red). The dashed straight line corresponds to $\I=\Pi_{\text s}/(2 \kappa_{\text {tot}})$.
\label{f:RndPlotMI_Raster}}
\end{figure}

\par
\subsection{Full comparison between irreversibility and correlations}
\begin{figure}[h!] 
\includegraphics[scale=.62]{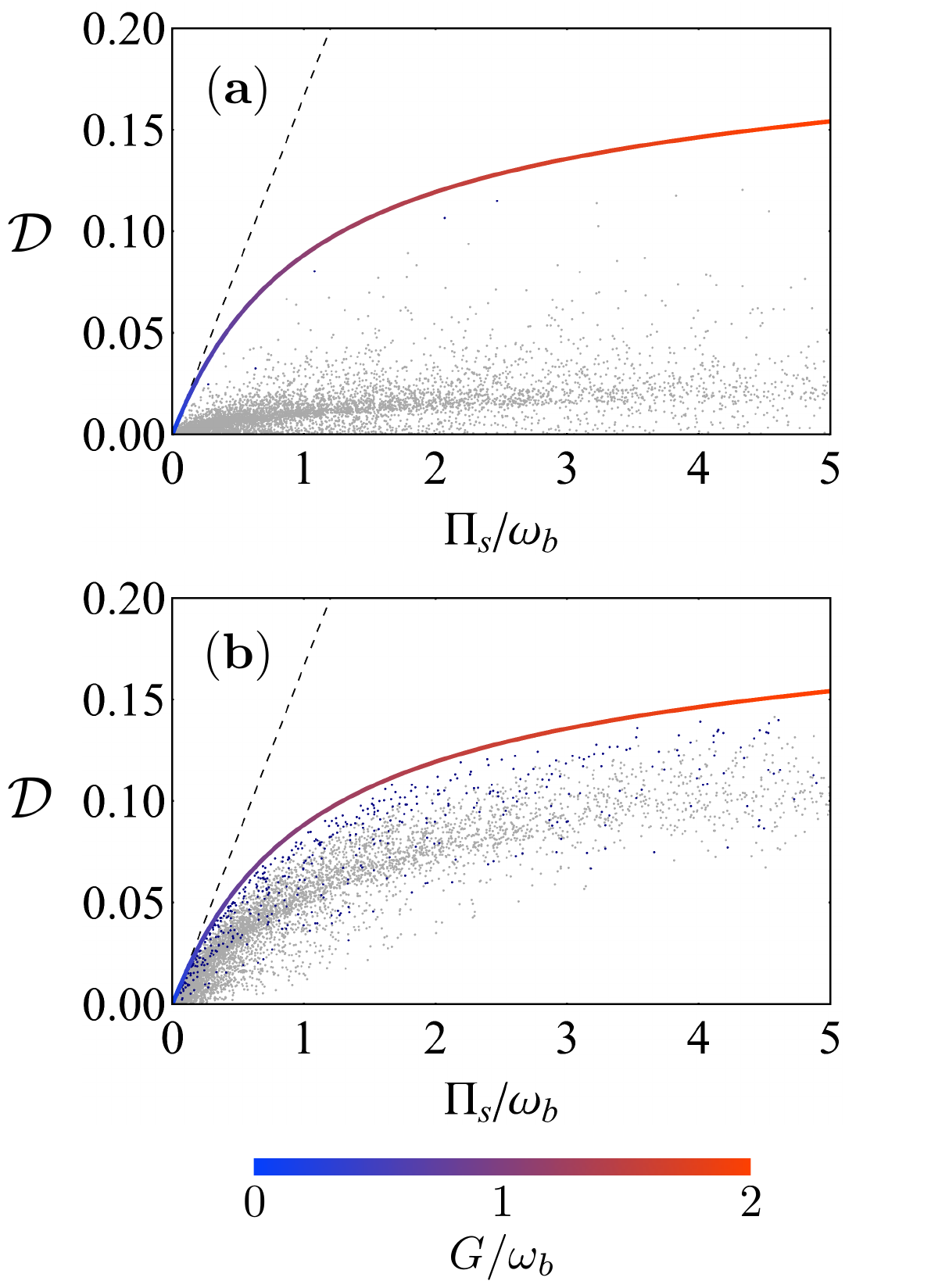}
\caption{Quantum discord $\D$ against the rescaled irreversible entropy production rate $\Pi_{\text s}/\omega_b$ for randomly generated states. Detail of the generation are the same as
in Fig. \ref{f:RndPlotMI_Raster}. The dashed straight line corresponds to $\D=\Pi_{\text s}/(4 \kappa_{\text {tot}})$.
\label{f:RndPlotDisc_Raster}}
\end{figure}

To better understand the relationship between $\Pi_{\text s}$ and the correlations $\I$ and $\D$, we proceed by randomly generating non-equilibrium steady states and 
evaluating the quantities of interest. The resulting plots are shown in Figs. \ref{f:RndPlotMI_Raster} and \ref{f:RndPlotDisc_Raster}, where we have considered uniform 
distributions for the dimensionless frequency $\omega_a\in [0,3]$, coupling values $G\in [0,2]$, and either $N_{a,b}\in [0,10]$, for the case shown in panel \textbf{(a)}, 
or $N_{a,b}\in [0,1]$ in \textbf{(b)}. 
The value of the dissipation rates has been kept fixed and set to $\kappa_a=0.5$ and $\kappa_b=1$. Furthermore, points corresponding to separable states are shown in gray, 
while entangled states are marked in blue. From Fig. \ref{f:RndPlotMI_Raster}, which compares $\Pi_{\text s}$ and $\I$, we see that all the randomly-generated points lie between two 
monotonically non-decreasing curves. Equivalently stated, for a given degree of irreversibility $\Pi_{\text s}$, {\it the total amount of correlations is both upper and lower bounded}. 
The upper bound turns out to be attained by oscillators having the same frequency and placed in contact with baths at the same temperature. 
For such a maximally symmetric configuration for analytical expressions of both $\Pi_{\text s}$ and $\I$ (the latter only for $\kappa_a=\kappa_b$) have been provided in 
Eq.~\eqref{PI_K1K2} and Eq.~\eqref{MI_Kappa}, respectively. On the other hand, the lower bound corresponds to states having $\omega_a=0$ and maximum imbalance in the 
initial populations, with all the excitations in the mode $\hat b$, i.e., $N_a=0$ and $N_b=N_{\text{max}}$ [with $N_{\text{max}}$=10 panel \textbf{(a)} and $N_{\text{max}}=1$ in 
\textbf{(b)}]. The class of states saturating the lower bound is therefore highly non symmetric, with the oscillator $\hat a$ freezed in the 
stationary state and its bath in the ground state.  
\par
In Fig. \ref{f:RndPlotMI_Raster} parametric curves with respect to the coupling $G$ are shown, for both the upper and the lower bounds, which go from $G_{\text{min}}=0$ (blue), to 
a maximum value $G_{\text{max}}=2$ (red).  The linear relation expressed in Eq.~\eqref{MainResult} is stressed by a dashed line, and we can see that is only accurate in 
a neighborhood of the origin of the plane where typically states with small coupling lie.
In particular, it can be checked that the configuration achieving the lower bound is stable for any value of the coupling $G$ and achieves an asymptote for high values of $\Pi_{\text s}$, 
which corresponds to the limit of infinite coupling. This value can be computed analytically and reads 
\begin{equation}
\I_{\text{min}}^{\infty}=\frac12 \log\left[\frac{(\kappa_a {+} 2 \kappa_b) (2 \kappa_a^3 {+} \kappa_b {+} 8 \kappa_a^2 \kappa_b {+} 8 \kappa_a \kappa_b^2 {+} \kappa_b^3)}
{\kappa_a\kappa_b(1 + \kappa_b^2)}\right].
\end{equation}
\par
In Fig. \ref{f:RndPlotMI_Raster} \textbf{(b)} we have considered lower values of the maximum number of initial thermal excitations allowed. While as we know this does not affect 
the upper bound, the separable/entangled nature of the stationary states gets drastically affected, with now many more entangled states generated by the dissipative process. 
We also notice that entangled states are denser in the region close to the upper bound. This fact is somehow expected since highly quantum-correlated states typically have high 
values of mutual information (classical plus quantum correlations).
%

In Fig. \ref{f:RndPlotDisc_Raster} we present a similar analysis for the R\'enyi-2 discord. The dashed straight line corresponds to $\D=\Pi_{\text s}/(4 \kappa_{\text {tot}})$ and, as
explained above, accurately describes states with $N_a=N_b=0$ and vanishing values of the coupling. In the small-coupling limit, it can be verified that the optimal measurement to be 
implemented on the mode $\hat b$ (the one that saturates  Eq.~\eqref{DG} and yields the quantum discord) is heterodyne detection (see Appendix D). On the other hand, the upper 
bound turns out to be attained by states having $\omega_a=0$ and $N_a=N_b=0$, i.e. with both the oscillators initially in the ground state. 
On the other hand, whenever $N_b=0$, homodyne detection turns out to be the optimal measurement. 

\section{Irreversibility and correlations in an Optomechanical System}\label{s:OptoCorr}
In this Section we particularize our findings to characterize irreversibility in a quantum optomechanical system. Optomechanical-like devices allow for the 
quantum-coherent manipulation of mesoscopic mechanical systems~\cite{Schliesser,Chan,Teufel,Groeblacher2}.  As such, they have recently attracted 
some attention as a platform where to investigate thermodynamics in the quantum regime and as a well-versed candidate to realize quantum thermal 
machines~\cite{Meystre, Myself, Elouard, Giovannetti, Kurizki, AbsRefr}. 
\par
The system we consider is an optomechanical cavity of length $L$ externally pumped by a laser having frequency $\omega_0$ and strength $\mathcal{E}=
\sqrt{2 P \kappa/ \hbar \omega_0}$, where $P$ is the incident laser power and $\kappa$ is the cavity decay rate. A mechanical mode of frequency $\omega_m$ 
is coupled to the cavity mode due to radiation-pressure interaction. The Hamiltonian of the system (in a rotating frame at the frequency of the external pump) 
reads~\cite{Aspelmeyer}
\begin{equation}\label{HamOpto}
\hat{H}_0=\hbar\Delta_0\hat{a}^{\dagger}\hat{a}+\frac{\hbar \omega_m}{2} \bigl(\hat{q}^2+\hat{p}^2\bigr) - \hbar g_0 \hat{a}^{\dagger}\hat{a} \hat{q}
+i \hbar\mathcal{E}(\hat{a}^{\dagger}-\hat{a}) \, ,
\end{equation}
where $\Delta_0=\omega_c-\omega_0$ is the cavity-pump detuning, $\hat q$ and $\hat p$ are the mechanical dimensionless quadratures, and the strength 
of the radiation pressure interaction is quantified by the single-photon coupling rate $g_0=\frac{\omega_c x_\mathrm{zpf}}{L}$, where $x_\mathrm{zpf}=
\sqrt{\hbar/ m \omega_m}$ is the zero-point term of the mechanical resonator. The dynamics of the system is also influenced by the presence of the environment. 
The mechanical mode is affected by the zero-mean Brownian stochastic force  $\hat{\xi}(t)$, which for high mechanical quality factors can be taken to be Markovian, 
and experiences dissipation at a rate $\gamma_m$. The intra-cavity field couples to the outside vacuum electromagnetic field, described by delta-correlated input noise 
$\hat a^{\text{in}}$. The noise vector entering the equations of motion is given by $\hat{N}=(0,\hat{\xi},\sqrt{2\kappa}\hat{X}^{\text{in}},\sqrt{2\kappa}\hat{Y}^{\text{in}})^T$.
For a strong driving, provided that the system remains in a stable regime, the cavity field reaches a steady value of the amplitude $\langle \hat a\rangle_{\text s}=\alpha$, and 
one can consider small quantum fluctuations around the classical steady state. This procedure leads to a linearized interaction $\hat H_I=2\hbar g \delta\hat{q}\delta\hat{X}$ 
between the mechanical $(\delta\hat{q})$ and optical $(\delta\hat{X})$ fluctuation operators and an enhanced optomechanical coupling 
$g=\frac{\sqrt{2}g_0\vert \mathcal{E}\vert}{\sqrt{\kappa^2+\Delta^2}}$, where $\Delta=\Delta_0-\frac{g_0^2\vert \alpha\vert^2}{\omega_m}$~\cite{Paternostro2006}. 
Therefore, the driven-dissipative optomechanical system matches our abstract model of linearly coupled quantum oscillators, provided that the oscillator $\hat a$ is identified  
with the quantum fluctuation of cavity field, and $\hat b$ with the annihilation operator of the mechanical fluctuation. Formally the correspondence is realized by replacing the vector 
$\hat u=(\hat q_a,\hat p_a,\hat q_b,\hat  p_b)^T$  by the vector of zero-mean fluctuation operators $(\delta\hat{q},\delta\hat{p},\delta\hat{X},\delta\hat{Y})^T$ and by
the following identification: $\omega_a=\Delta$, $\omega_b=\omega_m$, $G=2g$, $\kappa_a=\kappa$, $\kappa_b=\gamma_m$, $N_b=N$ and $N_a=0$. 
In the following  analysis we will employ dimensionless quantities. All the frequencies involved are understood to be in units of the mechanical frequency $\omega_m$, 
except for the captions of the figures. 
\begin{figure}[t!] 
\includegraphics[scale=.58]{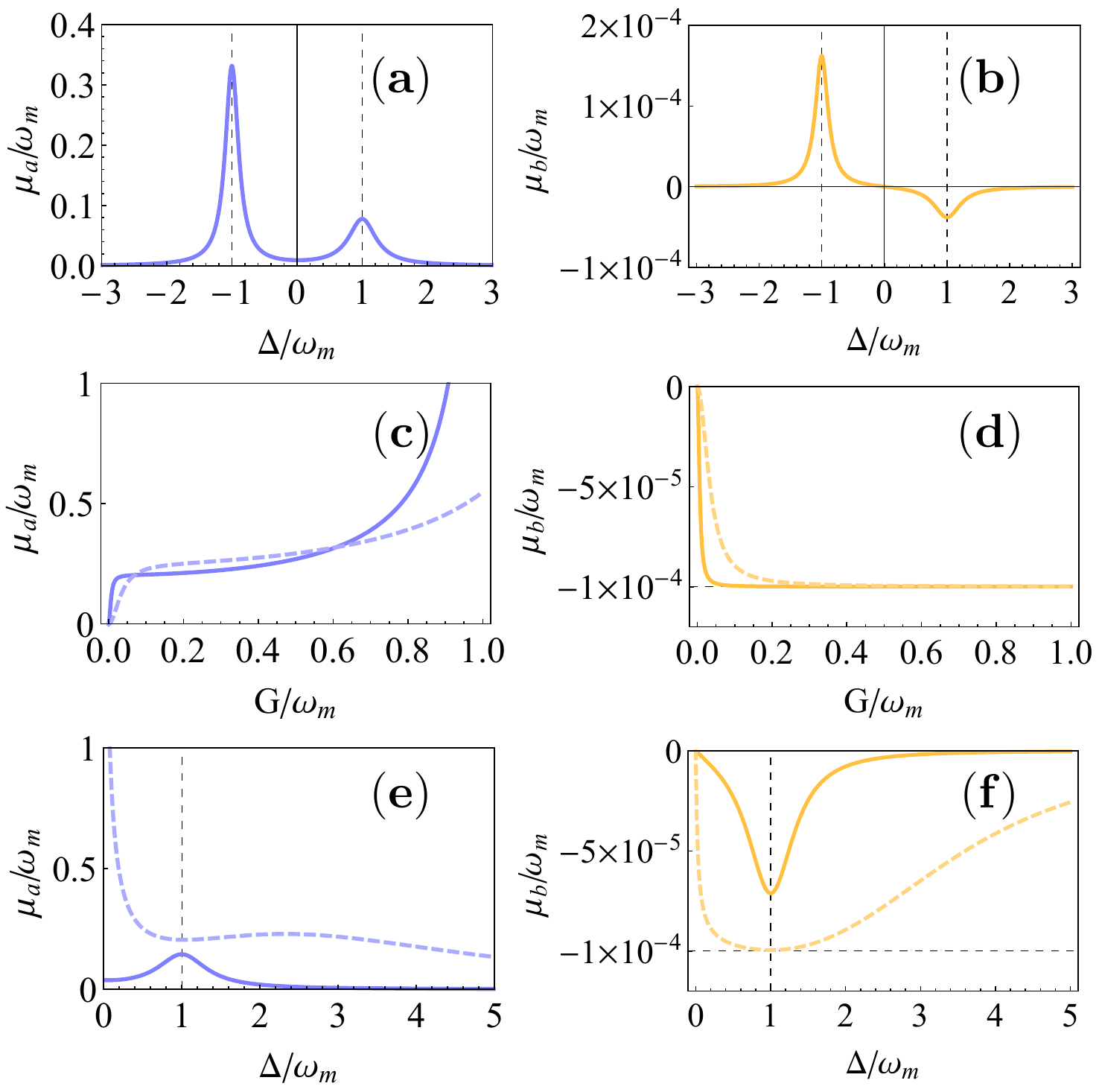}
\caption{Top row: Rescaled optical \textbf{(a)} and mechanical \textbf{(b)} contributions to the entropy production rate against the detuning $\Delta/\omega_m$, for $g=0.005
\omega_m$ $\kappa=0.2\omega_m$, $\gamma_m=10^{-4}\omega_m$ and $N=10^3$.  Middle row: optical \textbf{(c)} and mechanical \textbf{(d)} contributions against the 
rescaled coupling $g/\omega_m$; solid curves correspond to $\Delta=\omega_m$, while dashed ones to $\Delta=2\omega_m$. Other parameters are $\kappa=0.2\omega_m$, 
$\gamma_m=10^{-4}\omega_m$ and $N=10^3$. Panels \textbf{(e)} and \textbf{(f)} are obtained for the same values as \textbf{(a)} and \textbf{(b)}, except for the coupling which 
is given by $g=0.01\omega_m$ (solid curves) and  $g=0.1\omega_m$ (dashed curves).
\label{f:PlotOpto}}
\end{figure}
\par
One main difference with respect to the generic case of Section~\ref{s:IrrOsc} is that the effective frequency of the optical oscillator can now be tuned, and this enables us to 
explore the range of negative detunings, which was previously forbidden. In Fig. \ref{f:PlotOpto} \textbf{(a)} and \textbf{(b)} we display  the behavior of the the optical and mechanical 
contributions to the entropy production $\mu_a$ and $\mu_b$ against the detuning, whence we see that for small values of the coupling the system is stable also in the blue-detuned 
region $\Delta<0$. For large values of the detuning $\vert \Delta\vert \gg1$ the two oscillators are effectively decoupled and each of them comes to equilibrium with its own bath, 
thus leading to vanishing $\Pi_{\text s}$, in agreement to what observed in Sec.~\ref{s:EntProd}. The chosen values of the parameters place the system in the resolved sideband regime. 
This is reflected in the behavior of the entropy production rate, with both contributions $\mu_a$ and $\mu_b$ clearly peaked at the two mechanical sidebands. For frequencies 
$\Delta\approx 1$ the dominant beam splitter interaction $\hat H_{I}\propto \delta\hat a^{\dagger} \delta\hat b+\delta\hat a\delta\hat b^{\dagger}$ determines an enhanced heat transport, 
necessarily accompanied by an entropy increase. From Fig. \ref{f:PlotOpto} \textbf{(a)} we see that the peak is even more pronounced in the amplification regime. Indeed, for $\Delta\approx -1$ 
the two-mode squeezing like interaction $\hat H_{I}\propto \delta\hat a^{\dagger} \delta\hat b^{\dagger}+\delta\hat a\delta\hat b$ prevails, causing an exponential growth of the energies stored 
in the oscillators and inducing strong correlations between the two (EPR-like entanglement, in the infinite energy limit). And from our analysis it follows that the entropy production must 
correspondingly increase. On the other hand, from Fig. \ref{f:PlotOpto} \textbf{(b)} we see that $\mu_b$  changes sign, which clearly  captures the heating/cooling of the mechanical resonator. 
We  can thus conclude that the behavior of $\Pi_{\text s}$ and its contributions gives a full thermodynamical account of both the amplification and cooling regimes.
\par
In Fig. \ref{f:PlotOpto} \textbf{(c)} and \textbf{(d)} $\mu_{a,b}$ are shown as a function of the coupling, for different values of $\Delta$. The Routh-Hurwitz criterion determines the 
limiting value of the coupling (for a given detuning) beyond which the system is no longer stable~\cite{Hurwitz}. We can see that $\mu_a$ is divergent when approaching that value, which overall 
makes $\Pi_s$ diverge. On the other hand, $\mu_b$ tends to an asymptotic value  which, depending on the detuning $\Delta$, is approached for smaller or bigger coupling strengths. As expected, 
for $\Delta=1$ [solid line in panel \textbf{(d)}] the optimal working point for cooling is attained, which is manifested by the fact that the limiting value of $\mu_b$ is approached for 
the the smallest value of the coupling. 
\par
Finally, in Fig. \ref{f:PlotOpto} \textbf{(e)} and \textbf{(f)} we plot $\mu_a$ and $\mu_b$ for higher values of the optomechanical coupling. The blue-detuned region is now excluded because the 
semiclassical steady state is no longer stable. $\mu_a$ and $\mu_b$ increase (in absolute value) for increasing values of the coupling, which accounts for the fact that the stronger the interaction
between the oscillators the more irreversible the corresponding stationary process. In particular for $g=0.1$ the multi-photon quantum cooperativity  is $\mathcal{C}=4 g^2/\kappa \gamma N=50$, 
which places the system deeply in the quantum regime. The main feature to be noticed in plot \textbf{(e)} is that the behavior of the entropy production changes changes qualitatively: for $g=0.01$ 
$\mu_a$  still displays a maximum at the mechanical sideband, while for $g=0.1$ we see a divergence close to resonance and a local minimum around $\Delta=1$. On the other hand in panel 
\textbf{(f)} we see that the minimum always occurs at $\Delta=1$, the only difference being in its depth and its width. 
\par
\begin{figure}[h!] 
\centering
\includegraphics[scale=.51]{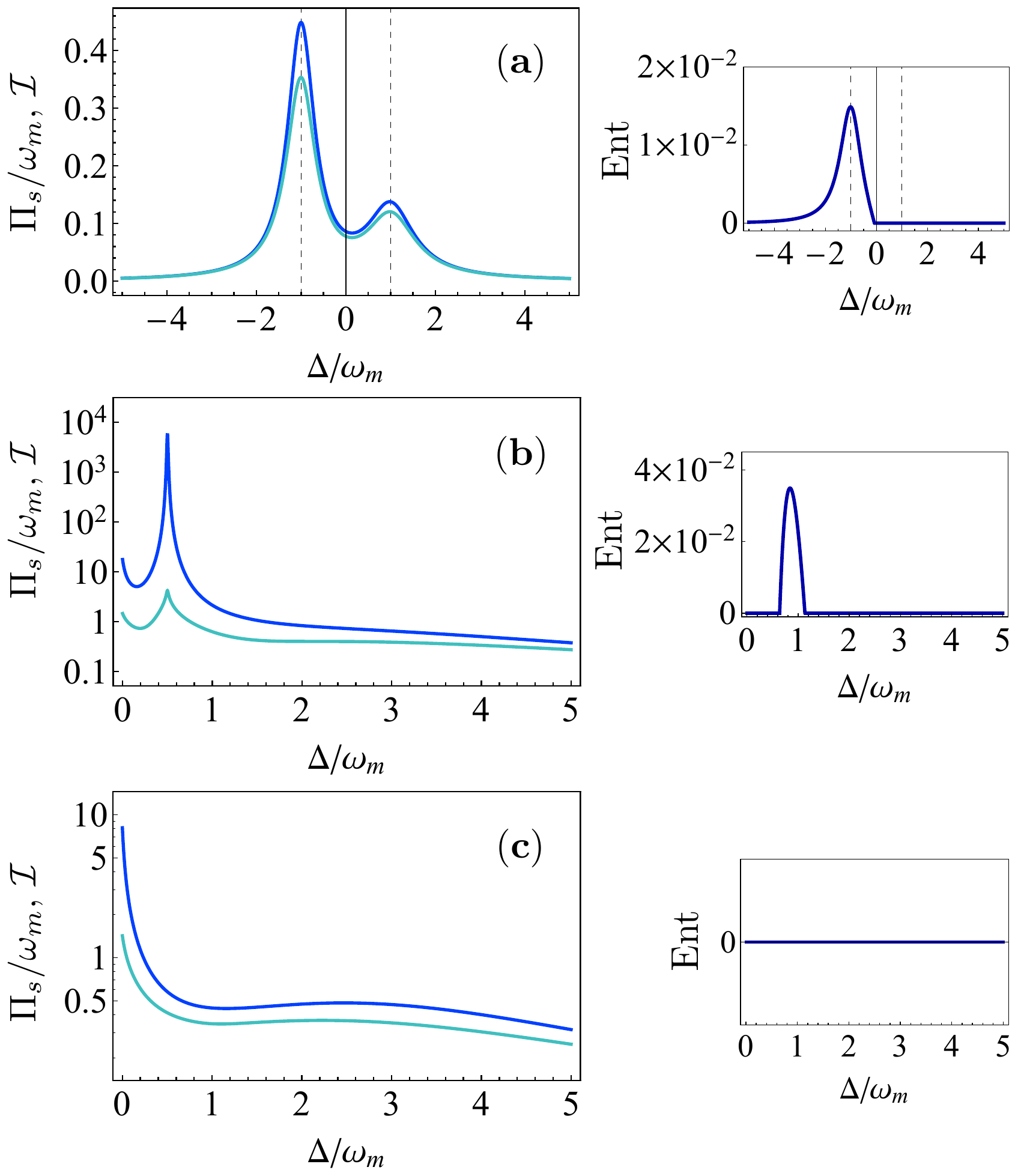}
\caption{Comparison between the steady-state entropy productions rate $\Pi_{\text s}$ (blue) and mutual information $\I$ (green water) as a function of 
the rescaled detuning $\Delta/\omega_m$ (left) together with the entanglement between the mechanical and optical field (right) quantified by the logarithmic 
negativity. \textbf{(a)} is for $g=0.1\omega_m$, $\gamma_m=10^{-2}\omega_m$ $\kappa=0.5\omega_m$ and $N=10$; in \textbf{(b)} logarithmic plot with same
parameters but strong coupling $g=\omega_m$. Finally in \textbf{(c)} same parameters as \textbf{(a)} except for $\gamma_m=10^{-4}\omega_m$ and $N=10^3$. 
\label{f:PlotOptoCorr}}
\end{figure}
\par
In Fig. \ref{f:PlotOptoCorr} we compare the entropy production rate $\Pi_{\text s}$ to the correlations established by the optomechanical interaction, as quantified by 
the mutual information $\I$. In panel \textbf{(a)} we have considered a small coupling ($g=0.1$) and a very low initial occupation of the mechanical oscillator ($N=10$). 
The two figures of merit are extremely close to each other for the whole range of detunings. Indeed, in this limit Eq.~\eqref{MainResult} safely applies.
The two-mode squeezing, which prevails in the region $\Delta \approx -1$, is an explicitly entangling operation that establishes strong quantum correlations, thus leading
to a sudden increase of both $\I$  and $\Pi_{\text s}$ that are strongly peaked at the mechanical sideband.
This is further confirmed by the presence of steady-state entanglement (quantified by the logarithmic negativity) in the blue detuned region, as shown in the right box.
In panel \textbf{(b)} the system is in the strong coupling regime, which causes t$\Pi_{\text s}$ to increase exponentially with respect to the total
amount of correlations. However, even if no analytical relation can be grasped in this case, from the plot is clear the similarity between the two figures
of merit persists. Finally, in \textbf{(c)} we consider more realistic values of the parameters $N=10^3$ $g=0.1$. Also for such a high imbalance in the number of excitations 
it is possible to see that $\I$  and $\Pi_{\text s}$ bear the same features, in particular being both maximum at resonance ($\Delta=0$).
\par
Expanding up to second order in $G$ and neglecting $\gamma$ we obtain the following expression for the entropy production rate
\begin{equation}
\Pi_{\text s}=\frac{2 \kappa G^2}{2 N+1}  \left(\frac{N^2}{(\Delta -1)^2+\kappa ^2}+\frac{(N+1)^2}{(\Delta +1)^2+\kappa ^2}\right) + \mathcal{O}(g^4)\, .
\end{equation}

\section{Conclusions}\label{s:Conclusions}

With quantum-limited control extending over systems of increasing size and complexity, the development of a dynamical theory of irreversible heat exchange 
for mesoscopic quantum systems undergoing finite-time transformation is fully due. In this work we have taken a step in this direction, by quantitatively 
assessing the irreversible entropy generated in an interacting quantum system by a stationary dissipative process. First, we presented two alternative expressions
for the stationary entropy production rate which relate in a simple way the irreversibility of the process to the features of the open system dynamics. 
Second, we established a connection between the entropy production rate and the amount of total and quantum correlations shared by the two subsystems. 
In order to do so, we found instrumental to quantify correlations by means of the the R\'enyi-2 entropy. We showed how the onset of irreversibility, which quantifies 
the departure from quasi-static reversible transformations, and the amount of correlations established by a process must be view as complementary features.  
Our results provide, for instance, a clear picture of the irreversibility taking place in a cavity optomechanical system. There, the entropy production rate turns out 
to be a very informative figure of merit, accounting for both sideband cooling and amplification regimes.
\par  
Possible applications of our work concern the optimization of quantum thermal machines operating at the steady state. For example, the model we considered can 
be taken as an instance of an autonomous quantum thermal machine that generates steady state quantum correlations being only powered by heat~\cite{Huber}, 
in which case our study accounts for the performances of such a machine. Furthermore, we plan to extend our analysis 
to a time-dependent scenario. This would enable to incorporate relaxation towards equilibrium as well as time-dependent driving and feedback, with the goal of
engineering genuinely out-of-equilibrium transformations in interacting quantum systems with the minimal generation of entropy.

\noindent
\section*{Acknowledgements}
This work was supported by the UK EPSRC (EP/L005026/1 and EP/J009776/1), the John Templeton Foundation 
(grant ID 43467) and the EU Collaborative Project TherMiQ (Grant Agreement 618074). Part of this work was supported by COST Action 
MP1209 ``Thermodynamics in the quantum regime''.


\renewcommand{\bibnumfmt}[1]{[A#1]}
\renewcommand{\citenumfont}[1]{A#1}

\section*{Appendix}

\subsection*{Appendix A: Quantifying information by the R\'enyi-2 entropy}
In this Appendix we motivate the adoption of correlation measures based on the R\'enyi-2 entropy. Let us consider a $n$-mode bosonic state $\varrho$ and let 
$\W_{\varrho}(X)$ be the Wigner function  associated to it, with $X \in \mathbb{R}^{2n}$. If $\W_{\varrho}(X)$ is a legitimate probability distribution we can 
compute its Shannon entropy $S(\W_{\varrho})=-2^{-n}\int_{\mathbb{R}^{2n}}\text{d}^{2n}X \W_{\varrho}(X) \log \W_{\varrho}(X)$. In particular, if the state is a 
(zero-mean) Gaussian state with 
covariance matrix $\sigma$, namely a state whose Wigner function has the form
\begin{equation}\label{WignerGauss}
\W_{\sigma}(X)=\frac{1}{\pi^n\sqrt{\det \sigma}}e^{-\frac12 X^T \sigma^{-1}X} \, ,
\end{equation}
its Shannon entropy $S(\W_{\sigma})\equiv S_{\sigma}$ can be explicitly evaluated and is given by 
\begin{equation}\label{ShannonWigner}
S_{\sigma}=\frac12 \log \det \sigma + n\log \pi e \, .
\end{equation}
Interestingly, for  Gaussian states this expression coincides, up to an additive constant, with the generalized R\'enyi entropy of order 2. The R\'enyi-$\alpha$ entropy is
defined as $S_{\alpha}(\varrho)=(1-\alpha)^{-1}\log \tr{\varrho^\alpha}$ with $\alpha\ge0$, and is a generalization of the usual entropy functional (one can indeed show 
that in the limit $\alpha \rightarrow1$ the von Neumann entropy $S_1=-\text{Tr} \varrho\log\varrho$ is recovered). When $\alpha=2$ the R\'enyi entropy takes the 
particularly simple expression $S_2(\varrho)=-\log \text{Tr} \varrho^2$, namely minus the logarithm of the purity of the state $\varrho$. For Gaussian states the purity 
can be easily evaluated as $\text{Tr} \varrho^2=({\pi/2})^{n}\int_{\mathbb{R}^{2n}}\text{d}^{2n}X \,\W_{\sigma}^2(X)=2^{-n}(\det \sigma)^{-1/2}$, yielding 
$S_2=\frac12 \log \det \sigma +n\log 2$. By comparing the latter expression with Eq.~\eqref{ShannonWigner} we thus have
\begin{equation}\label{TwoEntropies}
S_{\sigma}=S_2 + n\log \frac{\pi}{2} e \, ,
\end{equation}
which proves our statement.
\par
In Ref.~\cite{AAdesso} the strong subadditivity of $S_2$ for Gaussian states has been proved, and measures of both the mutual information ($\I_2$) and the quantum 
discord ($\D_2$) built upon it have been introduced. We must stress that the adoption of correlation measures based on the Von Neumann entropy  $S_1$ is perfectly 
legitimate, and actually standardly employed; for bipartite Gaussian states analytical expressions of both the mutual information ($\I_1$) and the quantum discord ($\D_1$)
are available~\cite{AIlluminati,ADiscordMatteo, ADiscordGerardo}. However, while these require the knowledge of the full symplectic spectrum, those based on the 
The R\'enyi-2 entropy only involve the local and global purities of the system, thus leading to much easier expressions. In our analysis we have adopted $\I_2$ and $\D_2$ 
to quantify the amount of total and quantum correlations, respectively. However, in order to ascertain the consistency of our choice, we compared $\I_2$ ($\D_2$) 
and $\I_1$ ($\D_1$) for the coupled oscillators in the stationary state, finding that they display the very same behavior, with  $\I_2$ ($\D_2$) in general 
assuming slightly smaller values than $\I_1$ ($\D_1$). 

\subsection*{Appendix B: Derivation of the stationary entropy production rate Eq.~\eqref{EntropyNESS}}
In this Appendix we provide explicit expressions for the entropy production rate $\Pi(t)$ and the entropy flux $\Phi(t)$ appearing in Eq.~\eqref{RateEq}.
In particular, we derive the expression for steady-state entropy production rate $\Pi_{\text s }=-\Phi_{\text s }$ shown in Eq.~\eqref{EntropyNESS}. The 
entropy is quantified by means of the Shannon entropy of the Wigner distribution of the state Eq.~\eqref{ShannonWigner}. Since we are interested in entropy rates, 
the additive constant in Eq.~\eqref{TwoEntropies} plays no role, so that we can remove the subscripts and denote the entropy just by $S$. The following derivation is 
based on Ref.~\cite{ALandi}, where however classical stochastic processes were considered.
\par
The dynamics of the system, as introduced in Sec.~\ref{s:IrrOsc}, has been described as the solution of the quantum Langevin equations for the vector 
$\hat u=(\hat q_a,\hat p_a,\hat q_b,\hat  p_b)^T$. It can be equivalently expressed in terms of a Fokker-Planck equation for the Wigner function $\W(u,t)$ 
\begin{equation}\label{FokkerPlanck} 
\partial_t \W=-\text{div} J(u,t) \, ,
\end{equation}  
where $u=( q_a, p_a, q_b,  p_b)^T$ is a point in the phase space and we have introduced the probability current vector  
\begin{equation}\label{J}
J(u,t)=A u \W(u,t)-\frac12 D \partial_u \W(u,t) \, .
\end{equation}
The drift and diffusion matrices $A$ and $D$ appearing in Eq.~\eqref{J} are defined in Sec.~\ref{s:IrrOsc}, and by $\partial_u$ we mean the phase-space gradient. 
By introducing the time-reversal operator $E=\text{diag}(1,-1,1,-1)$, the dynamical variables can be split according to their symmetry. The drift matrix $A$ is accordingly 
divided in an irreversible component $\Airr(E u)=E(\Airr u)$ which is even with respect to time reversal, and a reversible one $\Arev(E u)=-E(\Arev u)$, which id odd
~\cite{ASpinneyFord}. They can be constructed as  $\Airr=\frac12(A+EAE^T)$ and $\Arev=\frac12(A-EAE^T)$ and explicitly read 
\begin{equation}\label{Airr}
A^{\text{irr}}=\text{diag}\left(-\kappa_a,-\kappa_a,-\kappa_b,-\kappa_b\right) \, ,
\end{equation}
and
\begin{equation}\label{Arev}
A^{\text{rev}} =
\begin{pmatrix}
0 & \omega_a& 0 & 0 \\
-\omega_a& 0& G & 0 \\
0 & 0& 0 & \omega_b \\
G & 0& -\omega_b & 0 \\
\end{pmatrix}\,,
\end{equation}
while $D\equiv D^{\text{irr}}$. This separation induces a similar splitting in  the probability current  $J(u,t)=\Jrev(u,t)+\Jirr(u,t)$, where 
\begin{equation}\label{Jrev}
\Jrev(u,t)= \Arev u\W(u,t)\, , 
\end{equation}
and
\begin{equation}\label{Jirr}
\Jirr(u,t)= \Airr u\W(u,t)-\frac12 D \partial_u \W(u,t) \, .
\end{equation}
We notice that, given the form of Eq.~\eqref{Arev}, the reversible part of the probability current is divergence-less, i.e., $\text{div}\Jrev(u,t)= 
\W(u,t) \tr{ \partial_u(\Arev u)}=0$.
\par
The detailed balance condition is retrieved by demanding $J_{\text{eq}}(u)\equiv0$ and noting that the corresponding distribution $\W_{\text{eq}}(u)$ is
left unchanged by time-reversal, i.e.  $\W_{\text{eq}}(Eu)=\W_{\text{eq}}(u)$. Together these conditions correspond to microscopic reversibility and in this 
case the system's covariance matrix reduces to $\sigma_{\text{eq}}=(N_a+1/2)\mathds{1}_a\oplus (N_b+1/2)\mathds{1}_b$, thus describing two locally thermal
states. On the other hand, when $\text{div}J_{\text{s}}(u)\equiv0$ the system is found in a non-equilibrium stationary state. The two oscillators are now interacting
with each other and detailed balance is broken because of the enforcement of non-equilibrium boundary conditions. 
\par
If we substitute Eq.~\eqref{J} in the expression of the entropy rate we get  
\begin{equation}
\frac{\text{d}S}{\text{d} t}=\frac14 \int \text{d}u\, \text{div}J(u,t) \log \mathcal{W}(u,t) \, .
\end{equation}
Then, invoking the divergence-less of $\Jrev$ we can write the integrand as $\text{div}J(u,t) \log \W(u,t)=\text{div}\Jirr(u,t) \log \W(u,t)=\text{div}(\Jirr(u,t) 
\log \W(u,t))-\Jirr(u,t)^T \partial_u(\log \W(u,t))$ and we notice that the first term, when integrated, vanishes as a consequence of the Stokes theorem (we assume 
the probability density to be vanishing at the phase-space boundary). Therefore, the entropy rate is  expressed as
\begin{equation}
\frac{\text{d}S}{\text{d} t}=-\frac14 \int \text{d}u \frac{1}{\W(u,t)} \Jirr(u,t)^T\partial_u \W(u,t) \, .
\end{equation}
Finally, by plugging in Eq.~\eqref{Jirr} we recover the splitting shown in Eq.~\eqref{RateEq}
between entropy flux
\begin{equation}\label{Phi}
\Phi(t)=-\frac14 \int \text{d}u \, 2\Jirr(u,t)^T D^{-1}\Airr u \, ,
\end{equation}
and entropy production rate 
\begin{equation}\label{PiRate}
\Pi(t)=\frac14 \int  \text{d}u \frac{2}{\mathcal{W}(u,t)}\Jirr(u,t)^T D^{-1}\Jirr(u,t) \, .
\end{equation}
It has been possible to identify the last expression with the entropy production rate by virtue of its non-negativity, being Eq.~\eqref{PiRate} the 
integral of  a quadratic form. Moreover, for a Gaussian state it is easy to show that the following relation holds, $\partial_u\mathcal{W}_{\sigma}(u,t)
=-\mathcal{W}_{\sigma}(u,t)\sigma^{-1}(t)u$, so that irreversible component of probability current Eq.~\eqref{Jirr} becomes 
$J^{\text{irr}}(u,t)=\mathcal{W}_{\sigma}(u,t)(\Airr u+D)\sigma^{-1}(t)u$ and the integral in Eq.~\eqref{PiRate} can be explicitly carried 
out, giving 
\begin{equation}\label{PiExplicit}
\Pi(t)=\frac12\tr{\sigma^{-1}D} + 2\tr{\Airr}+2\tr{(\Airr)^TD^{-1}\Airr\sigma} \, .
\end{equation}
\par
On the other hand, if we take the derivative of Shannon entropy of the Wigner function Eq.~\eqref{WignerGauss} we get $\frac{\text{d}S}{\text{d} t}
=\frac12\tr{\sigma^{-1}\partial_t\sigma}$, where we applied the Jacobi formula to the derivative of the determinant. By inserting the evolution of the 
covariance matrix Eq.~\eqref{SigmaEqMotion} we get
\begin{equation}
\frac{\text{d}S}{\text{d} t}=\frac12\tr{\sigma^{-1}(t)D} + \tr{\Airr} \, .
\end{equation}
By comparing the latter expression with Eq.~\eqref{PiExplicit} we can thus write
\begin{equation}
\Pi(t)=\frac{\text{d}S}{\text{d} t}-\Phi(t)
\end{equation}
where
\begin{equation}
\Phi(t)=-\tr{\Airr}-2\tr{(\Airr)^TD^{-1}\Airr\sigma(t)}\,.
\end{equation}
In particular, when the system attains a stationary state we have $\Pi_\text{s}=-\Phi_\text{s}$, whose expression is given by
\begin{equation}\label{Pi_s}
\Pi_\text{s}=\tr{\Airr}+2\tr{(\Airr)^TD^{-1}\Airr\sigma_\text{s}}\,.
\end{equation}
The latter is now in matrix form and the trace can be easily evaluated, giving
\begin{equation}\label{Pi_s}
\Pi_\text{s}= 2 \kappa_a \left( \frac{[\sigma_\text{s}]_{11}+[\sigma_\text{s}]_{22}}{2N_{a} +1}-1\right)   + 
2 \kappa_b \left( \frac{[\sigma_\text{s}]_{33}+[\sigma_\text{s}]_{44}}{2N_{b} +1}-1\right) \,
\end{equation}
which is the result displayed  in Eq.~\eqref{EntropyNESS}. 
\par

\subsection*{Appendix C: Derivation of the equivalent expression Eq.~\eqref{Mu_nondiag}}

In this Appendix we briefly show how to derive the expression for the  stationary entropy production rate shown in   
Eq.~\eqref{Mu_nondiag}. As shown in  Sec.~\ref{s:IrrOsc}, the steady-state covariance matrix obeys the Lyapunov 
equation
\begin{equation}
A \sigma_{\text s} + \sigma_{\text s} A^T = - D \, ,
\end{equation}
here rewritten for the sake of convenience, which imposes constraints between the elements of $ \sigma_{\text s}$.
Explicitly, the diagonal elements relative to the mode $\hat a$ are re-expressed as   
\begin{align}
[\sigma_\text{s}]_{11}&=\frac12 +N_a  +\frac{\omega_a}{\kappa_a} [\sigma_\text{s}]_{12}  \, , \\
[\sigma_\text{s}]_{22}&=\frac12+ N_a+ \frac{G[\sigma_\text{s}]_{23} -\omega_a[\sigma_\text{s}]_{12}}{\kappa_a} \, .
\end{align}
An analogous relation is found for the diagonal elements relative to the mode $\hat b$, upon the following substitutions
$[\sigma_\text{s}]_{11}\rightarrow[\sigma_\text{s}]_{33}$, $[\sigma_\text{s}]_{12}\rightarrow[\sigma_\text{s}]_{34}$, 
$[\sigma_\text{s}]_{23}\rightarrow[\sigma_\text{s}]_{14}$, and $a\rightarrow b$. By plugging those expressions into
 Eq.~\eqref{EntropyNESS}, the required expression for $\mu_{a,b}$ Eq.~\eqref{Mu_nondiag} is easily found.
 
\subsection*{Appendix D: R\'enyi-2 Gaussian discord}
In this Appendix we address more in detail the measure of discord based on the R\'enyi-2 entropy (here simply referred as $\D$), which has been used 
in the main text to quantify the amount of quantum correlations beyond entanglement in Gaussian states. In one Appendix of Ref.~\cite{AAdesso} an 
explicit expression of the R\'enyi-2 discord for Gaussian states in standard form has been provided. However, both for the sake of completeness and 
because we adopt a different convention, in the following we show how to derive such expression. Moreover, we extend it to the case of generic two-mode 
Gaussian states, which is the relevant case for our investigation, expressing $\D$ as a function of the symplectic invariants. The discord $\D$ between 
modes $\hat a$ and $\hat b$, when performing a measurement on $\hat b$, is defined as the following difference
\begin{equation}
\D(\sigma_{a\vert b})=\mathcal{I}(\sigma_{a:b})-\mathcal{J}(\sigma_{a\vert b}) \, .
\end{equation}
The first term on the right-hand side is the mutual information $\mathcal{I}(\sigma_{a:b})=\frac12 \log \left(\frac{\det \sigma_a\, \det \sigma_b}{\det \sigma_{ab}}\right)$, 
while the second $\mathcal{J}(\sigma_{a\vert b})=\sup_{\pi_b(X)}\{S(\sigma_a)-\int\text{d}X p_X S(\sigma_{a\vert X}^{\pi_b}) \}$ quantifies the one-way classical 
correlations when performing on $\hat b$ a measurement $\hat\pi_b(X)\ge0$, $\int \text{d}X\hat\pi_b(X)=\mathds{1}$. If we focus on 
Gaussian measurements, we can consider a POVM of the form $\hat\pi_b(X)=\pi^{-1}\hat D_b (X)\hat{\varrho}^{\pi_b}\hat D_b^{\dagger}(X)$ where 
$\hat D_b(X)=\exp(X \hat b^{\dagger}-X^* \hat b)$ is the displacement operator and $\hat{\varrho}^{\pi_b}$ a pure Gaussian state with covariance 
matrix $\gamma^{\pi_b}=R(\theta)\text{diag}(\lambda/2,\lambda^{-1}/2)R(\theta)^T$, where $\lambda \in[0,\infty]$ and $R(\theta)=\cos\theta \mathds{1}
-i\sin\theta \sigma_y$ is a rotation matrix. The conditional state of mode $\hat a$ once the measurement $\hat\pi_b(X)$ has been performed 
turns out to be independent on the outcome of the measurement, i.e. $\sigma^{\pi_b}_{a\vert X}\equiv \sigma_a^{\pi_b}$. Its expression is given by 
$\sigma_a^{\pi_b}=\sigma_a- c_{ab}(\sigma_b+\gamma^{\pi_b})^{-1}c_{ab}^T$, where the block-form of the covariance matrix has been defined in Eq.~\eqref{sigma}.  
We thus have $\int\text{d}X p_X S(\sigma_{a\vert X}^{\pi_b})=S(\sigma_a^{\pi_b})$ and the expression of one-way classical correlations reduces to
\begin{equation}
\mathcal{J}(\sigma_{a\vert b})=\sup_{\pi_b}\frac12 \log \left(\frac{\det \sigma_a}{\det \sigma_a^{\pi_b}}\right) \, .
\end{equation}
Accordingly, the R\'enyi-2 discord takes the  form
\begin{equation}\label{DiscMin}
\D(\sigma_{a\vert b})= \frac12 \log (\det \sigma_b)-\frac12 \log (\det \sigma_{ab})+\inf_{\pi_b}\frac12 \log (\det \sigma_a^{\pi_b}) \, .
\end{equation}
\par
Let us first consider Gaussian states in standard form, i.e. those whose covariance matrix is given by
\begin{equation}
\sigma_{\text{std}} =
\begin{pmatrix}
a & 0 & c & 0 \\
0 & a & 0 & d \\
c & 0 & b & 0 \\
0 & d & 0 & b \\
\end{pmatrix}\,.
\end{equation}
For such states the minimization in Eq.~\eqref{DiscMin} can be carried out analytically; if we denote $E_{a\vert b}^{\text{min}}=\inf_{\theta,\lambda}\det \sigma_a^{\pi_b}$,
and assuming $c>\vert d\vert$, explicit calculations show that its expression is given by
\begin{equation}\label{DiscPVM}
E_{a\vert b}^{\text{min}}=a \left(a-\frac{c^2}{b}\right)\, 
\end{equation}
whenever the condition $\left[4 a b^2 d^2-c^2 \left(a+4 b d^2\right)\right] \left[4 a b^2 c^2-d^2 \left(a+4 b c^2\right)\right]<0$, is satisfied and
\begin{widetext}
\begin{equation}\label{DiscPOVM}
E_{a\vert b}^{\text{min}}=\frac{a^2 \left(1-4 b^2\right)^2-4 a b \left(4 b^2-1\right) \left(c^2+d^2\right)+4 \left(\sqrt{c^2 d^2 \left(-4 a b^2+a+4 b c^2\right) 
\left(-4 a b^2+a+4 b d^2\right)}+\left(4 b^2+1\right) c^2 d^2\right)}{\left(1-4 b^2\right)^2} \, ,
\end{equation}
\end{widetext}
otherwise. It can be verified that the minimum in the first expression Eq.~\eqref{DiscPVM} is achieved by implementing homodyne detection on the mode $\hat b$,  
namely a projective measurement, e.g. along $\theta=0$ with an infinite amount of squeezing $\lambda\rightarrow0,\infty$. On the other hand, the minimum in the 
second expression of Eq.~\eqref{DiscPOVM} is achieved by a positive operator-valued measurement.

The steady state $\sigma_{\text s}$ of the dissipative dynamics Eq.~\eqref{SigmaEqMotion} is not in standard form. However, since the discord is invariant under local unitary 
operations, we can cast our state in standard form by means of a suitable set of local unitaries and then compute its discord. Equivalently, we can introduce the symplectic 
invariants $I_1=\det \sigma_a$, $I_2=\det \sigma_b$, $I_3=\det c_{ab}$ and $I_4=\det \sigma_{ab}$, and write the discord as a function of those. This leads us to the following 
expression for $E^{\mathrm{min}}_{a|b}$
\begin{widetext}
\begin{equation}\label{emin}
E^{\mathrm{min}}_{a|b}=
\left\{
\begin{array}{cc}
\frac{\left(I_1 I_2-I_3^2+I_4\right)-\sqrt{\Lambda_-^2 - 2 I_4 \Lambda_+ + I_4^2}}{2 I_2} & \mathrm{if} \; \; 
4I_4(\Gamma_+I_3^2-I_4)+I_1(\Gamma_+I_3^2+8 I_2 I_4)< 4I_1^2I_2^2,\\
 \\
\frac{I_1-4 I_1 I_2+ 4 \left(2 I_3^2-\Gamma_- I_4+\left| I_3\right|  \sqrt{I_1-4 (\Gamma_- I_4+ \Lambda_-)}\right)}{\Gamma_-^2} & \mathrm{otherwise},
\end{array}
\right.
\end{equation}
\end{widetext}
where we defined the quantities $\Gamma_{\pm}=1\pm4 I_2$, and  $\Lambda_{\pm}=I_1 I_2\pm I_3^2$. The expression of the R\'enyi-2 discord for a 
generic two-mode Gaussian state is thus given by
\begin{equation}\label{DiscFinal}
\D=\frac12 \log\left(\frac{I_2 \,E^{\mathrm{min}}_{a|b}}{I_4} \right) \, .
\end{equation}

\end{document}